\documentclass[twocolumn]{aastex631}

\usepackage{graphicx}	
\usepackage{amsmath}	
\usepackage{xfrac}
\usepackage{amssymb}	
\usepackage{mathtools}
\usepackage{newtxtext,newtxmath}
\usepackage{array}
\usepackage{float}
\usepackage{enumerate}
\usepackage[utf8]{inputenc}
\usepackage{ltablex} 
\keepXColumns        
\newcommand{\amaffil}{Citizen Science Contributor \footnote{A list of contributing observatories can be found in Appendix \ref{appendix:transit_data}.}}

\shortauthors{Hagey et al.}

\begin{document}

\title{TrES-1\,b: A Case Study in Detecting Secular Evolution of Exoplanet Orbits}

\correspondingauthor{Simone R. Hagey}
\email{shagey@phas.ubc.ca}

\author[0000-0001-8072-0590]{Simone R. Hagey}
\affiliation{Department of Physics and Astronomy, The University of British Columbia, 6224 Agricultural Rd. Vancouver, BC V6T 1Z1, Canada}

\author[0000-0002-5494-3237]{Billy Edwards}
\affiliation{SRON, Netherlands Institute for Space Research, Niels Bohrweg 4, NL-2333 CA, Leiden, The Netherlands}

\author[0000-0003-3840-1793]{Angelos Tsiaras}
\affiliation{Department of Physics, Aristotle University of Thessaloniki, 54124 Thessaloniki, Greece}

\author[0000-0002-0574-4418]{Aaron C. Boley}
\affiliation{Department of Physics and Astronomy, The University of British Columbia, 6224 Agricultural Rd. Vancouver, BC V6T 1Z1, Canada}

\author{Anastasia Kokori}
\affiliation{Department of Physics and Astronomy, University College London, Gower Street, London, WC1E 6BT, UK}

\author[0000-0001-8511-2981]{Norio Narita}
\affiliation{Komaba Institute for Science, The University of Tokyo, 3-8-1 Komaba, Meguro, Tokyo 153-8902, Japan}
\affiliation{Instituto de Astrofísica de Canarias (IAC), 38205 La Laguna, Tenerife, Spain}

\author{Pedro V. Sada}
\affiliation{Departamento de Física y Matemáticas, Universidad de Monterrey, Av. I. Morones Prieto 4500 Pte., 66238 San Pedro Garza García, N.L., México}

\author[0000-0003-2060-4912]{Filip Walter}
\affiliation{Observatory and Planetarium Prague, Stefanik Observatory, Strahovská 205, Praha, Czech Republic}
\affiliation{Czech Astronomical Society, Fričova 298 251 65 Ondrějov, Czech Republic}

\author{Robert T. Zellem}
\affiliation{Jet Propulsion Laboratory, California Institute of Technology, 4800 Oak Grove Dr., Pasadena, CA 91109, USA}

\author[0000-0001-7234-7167]{Napaporn A-thano}
\affiliation{\amaffil}
\affiliation{National Astronomical Research Institute of Thailand, 260 Moo 4, Donkaew, Mae Rim, Chiang Mai, 50180, Thailand}

\author{Kevin B. Alton}
\affiliation{\amaffil}
\affiliation{UnderOak Observatory, 70 Summit Ave, Cedar Knolls, NJ 07927, USA}

\author{Miguel Ángel Álava Amat}
\affiliation{\amaffil}

\author[0000-0001-6981-8722]{Paul Benni}
\affiliation{\amaffil}
\affiliation{Acton Sky Portal (private observatory), Acton, MA USA}

\author{Emmanuel Besson}
\affiliation{\amaffil}

\author{Patrick Brandebourg}
\affiliation{\amaffil}

\author{Marc Bretton}
\affiliation{\amaffil}
\affiliation{Observatoire des Baronnies Provençales, Route de Nyons, F-05150 Moydans, France}
\affiliation{LABSCAN, Laboratoire d'Astrophysique des Baronnies : Science Citoyenne et Action pour la Nuit}

\author{Mauro Caló}
\affiliation{\amaffil}

\author{Martin Valentine Crow}
\affiliation{\amaffil}

\author{Jean-Christophe Dalouzy}
\affiliation{\amaffil}

\author{Marc Deldem}
\affiliation{\amaffil}

\author[0000-0001-9948-0443]{Tõnis Eenmäe}
\affiliation{\amaffil}
\affiliation{Tartu Observatory of Tartu University, Observatooriumi 1, 61602 Tõravere, Estonia}

\author{Stephane Ferratfiat}
\affiliation{\amaffil}
\affiliation{Observatoire des Baronnies Provençales, Route de Nyons, F-05150 Moydans, France}

\author[0000-0002-4308-2339]{Pere Guerra}
\affiliation{\amaffil}
\affiliation{Observatori Astronòmic Albanyà, Camí de Bassegoda S/N, Albanyà E-17733, Girona, Spain}

\author{Gary Vander Haagen}
\affiliation{\amaffil}

\author{Ken Hose}
\affiliation{\amaffil}

\author{Adrian Jones}
\affiliation{\amaffil}

\author{Yves Jongen}
\affiliation{\amaffil}
\affiliation{Observatoire de Vaison-La-Romaine, Départementale 51, près du Centre Equestre au Palis, F-84110 Vaison-La-Romaine, France}

\author{Didier Laloum}
\affiliation{\amaffil}
\affiliation{AAVSO, 49 Bay State Rd., Cambridge, MA 02138, USA}

\author[0000-0003-1027-9853]{Stefano Lora}
\affiliation{\amaffil}
\affiliation{MarSEC (Marana Space Explorer Center), c/a Pasquali, Marana di Crespadoro VI, I-36070, Italy}

\author[0000-0003-3779-6762]{Alessandro Marchini}
\affiliation{University of Siena, Dept. of Physical Science, Earth and Environment, Astronomical Observatory, Via Roma 56, I-53100 Siena, Italy}

\author{Jacques Michelet}
\affiliation{\amaffil}

\author{Matej Mihelčič}
\affiliation{\amaffil}

\author{Johannes Mieglitz}
\affiliation{\amaffil}

\author{Eric Miny}
\affiliation{\amaffil}

\author{David Molina}
\affiliation{\amaffil}

\author[0000-0002-9920-434X]{Mario Morales Aimar}
\affiliation{\amaffil}
\affiliation{Observatorio de Sencelles, Sonfred Rd. 1, 07140 Sencelles, Balearic Islands, Spain}

\author{Raphael Nicollerat}
\affiliation{\amaffil}

\author[0000-0003-3833-2695]{Ivo Peretto}
\affiliation{\amaffil}
\affiliation{MarSEC (Marana Space Explorer Center), c/a Pasquali, Marana di Crespadoro VI, I-36070, Italy}

\author[0000-0002-2190-3319]{Manfred Raetz}
\affiliation{\amaffil}
\affiliation{Bundesdeutsche Arbeitsgemeinschaft für veränderliche Sterne e.V. (BAV), Germany}

\author{François Regembal}
\affiliation{\amaffil}

\author{Robert Roth}
\affiliation{\amaffil}

\author{Lionel Rousselot}
\affiliation{\amaffil}

\author[0009-0000-6809-7679]{Mark Salisbury}
\affiliation{\amaffil}

\author{Darryl Sergison}
\affiliation{\amaffil}

\author[0000-0002-6176-9847]{Anaël Wünsche}
\affiliation{\amaffil}
\affiliation{Observatoire des Baronnies Provençales, Route de Nyons, F-05150 Moydans, France}
\affiliation{LABSCAN, Laboratoire d'Astrophysique des Baronnies : Science Citoyenne et Action pour la Nuit}

\author{Jaroslav Trnka}
\affiliation{\amaffil}
\affiliation{Hvězdárna Slaný (Slaný Observatory), Nosačická 1713, Slaný, Czech Republic}
\affiliation{Czech Astronomical Society, Fričova 298 251 65 Ondrějov, Czech Republic}

\begin{abstract}
We present a comprehensive analysis of transit, eclipse, and radial velocity data of the hot Jupiter TrES-1\,b and confirm evidence of orbital variations on secular timescales. Apparent variations due to systemic motion and light travel time effects have been ruled out, indicating that the observed changes are dynamical in origin. Joint modeling of the TrES-1\,b data favors an apsidal precession model, but the rapid precession rate of $4^\circ$ yr$^{-1}$ cannot be explained without invoking an undetected close-in planetary companion, which remains unseen in the data. While radial velocity measurements reveal a previously undetected companion candidate on a wide, eccentric orbit, it is unlikely to drive the observed evolution of TrES-1\,b. However, an orbital decay model provides a plausible alternative if the loss of orbital energy is driven by planetary obliquity tides. We find that the best-fit orbital decay rate of $-7.1^{\,+1.5}_{-1.6}$ ms yr$^{-1}$ is aligned with theoretical predictions for modified tidal quality factors of hot Jupiters if TrES-1\,b has a planetary obliquity $\varepsilon_p > 30^\circ$. We encourage follow-up observations of this system, particularly of eclipse timing and radial velocities, to further constrain the nature of the observed evolution. This paper provides a practical framework for studying secular variations and aims to accelerate future research on similar systems. 
\end{abstract}

\section{Introduction} \label{sec:intro}
Planetary systems evolve with time, driven by multiple processes that include planet-planet and star-planet interactions. While this evolution is interesting in and of itself, it also provides a way to probe exoplanetary interiors and can help to resolve outstanding issues concerning the provenance of short-period planets. Hot Jupiters (HJs) are of particular interest because (1) they have tight orbits about their host stars, which amplify star-planet interactions, and (2) their deep transits and short periods enable precision observations with even modest facilities. Moreover, some HJs have been known for one to two decades, making it possible to detect, in principle, subtle orbital changes that require many epochs of observations. 

HJs may be particularly susceptible to a process known as orbital decay, a consequence of the transfer of angular momentum from the planet to the host star through tides, which slowly shrinks the planet's orbit, possibly leading to planetary engulfment by the star. Such orbital decay can be revealed by detecting a growing difference between the expected and measured transit midtimes, known as transit timing variations (TTVs), which arise from the decreasing planetary period as its orbit shrinks.\footnote{In the literature, TTV typically refers to short-term transit timing variations caused by gravitational interactions with additional planets, especially near mean-motion resonances. Since we are analyzing long-term changes in TrES-1\,b’s orbit over many years, we use the term ``secular'' TTVs to distinguish these trends from short-term variations.} This effect has been unambiguously observed for the HJ WASP-12\,b, for which the orbital period is changing at a rate of about $-30$ $\rm ms~yr^{-1}$ \citep[e.g.,][]{Patra2017,Yee2019,Turner2020, Wong2022}. 

However, these secular variations can be mimicked by another physical process, called apsidal precession, in which the argument of pericenter of a planet's orbit, $\omega_p$, changes gradually due to deviations from a Keplerian gravitational potential. Here, the line connecting the pericenter and apocenter of the orbit rotates through 360 degrees in one precession period at a rate that is expected to be on the order of a few degrees per year for the shortest-period HJs \citep{Ragozzine2009}. Distinguishing between apsidal precession and orbital decay is possible because the former causes periodic secular TTVs, while the latter does not. Nonetheless, if observations rely on transit midtimes only, then a large number of epochs will typically be needed to determine the underlying TTV mechanism.

Fortunately, the inclusion of secondary eclipses can greatly reduce the observational baselines needed to rule out one mechanism or the other. Moreover, radial velocity (RV) measurements add yet another lever for understanding a system's dynamics, while combining all three -- transits, eclipses, and RV -- can speed up the process of determining the cause of an HJ's evolving orbit, or perhaps detect it in the first place. For example, \citet{Alvarado2024} explored the use of radial velocity measurements to detect or constrain orbital decay.

Where should we look? This is not obvious. Numerous HJs have been proposed in the literature as prime candidates for decay due to their ultrashort orbital periods ($\lesssim$1 days), such as WASP-18\,b, WASP-19\,b, and WASP-43\,b. Yet, these systems show no convincing signs of decay. Despite this, possible long-term timing deviations have been detected in large samples \citep[e.g.,][]{Patra2020,Ivshina2022,Hagey2022, Yeh2024,Wang2024,Alvarado2024,Adams2024,Yalcinkaya2024}. Thus, we can turn the problem around and follow these ``leads'' with the hope that in-depth analyses can reveal whether the orbits are truly changing.\footnote{We direct the reader to \citet{Adams2024} for a thorough summary of current candidate systems.}

In 2021, as part of a study analyzing data from the Exoplanet Transit Database (ETD), we found evidence that HJ TrES-1\,b is deviating from a linear ephemeris at a constant rate of $-16.0 \pm 3.7$ ms yr$^{-1}$ \citep{Hagey2022}. Concurrently, \citet{Ivshina2022} published a transit timing study, which included TrES-1\,b, combining many literature observations with recent TESS data. They found the TrES-1\,b period to be changing at a rate $-18.36 \pm 3.73$ ms yr$^{-1}$. Furthermore, \citet{Adams2024} recently included TrES-1\,b in a broad analysis of HJ transit times, finding a best-fit period derivative of $-16.4 \pm 4.9$ ms yr$^{-1}$. The consistency across these independent studies highlights the need for a deeper investigation of TrES-1\,b. Each of these analyses relied solely on transit midtimes, as they were part of broader surveys covering many exoplanets. TrES-1\,b’s unique status as one of the first transiting planets discovered provides a wealth of archival data, including eclipse midtimes and RV measurements. Leveraging this well-rounded dataset, we present a focused study of TrES-1\,b, using the \href{https://github.com/simonehagey/orbdot}{\texttt{OrbDot}} Python package \citep{orbdot} for the analysis.

TrES-1\,b was discovered in 2004 \citep{Alonso2004}, the first of the Trans-Atlantic Exoplanet Survey (TrES). It is a $0.697$ M\textsubscript{J}, $1.067$ R\textsubscript{J} hot Jupiter on a $3.03$ day orbit \citep{torres_t1,Bonomo2017}. The host star, TrES-1, is a 0.878 M$_{\odot}$, 0.807 R$_{\odot}$ K0 V type star with a rotational period of $40.2 \pm 0.1$ days \citep{Dittmann2009} and solar-like metallicity \citep{torres_t1,Sozzetti2006}. The TrES-1 system was studied thoroughly in the years following its discovery and has been the subject of various studies seeking evidence of perturbing bodies in the system via short-term TTVs, but no significant variations of this type have been detected \citep{steffen_t1,Hrudkova2009,Raetz2009,Baluev2015,Rabus2008,rabus_t1_2009,Yeung2022}.

\begin{table}
  \addtocounter{table}{-1}
\centering
\begin{tabularx}{0.97\linewidth}{cccl}
    \textbf{Parameter} & \textbf{Unit} & \textbf{Value} & \textbf{Reference} \\ 
    \hline \hline \noalign{\vskip 0.5mm} 
    $M_\star$ & M$_{\odot}$ & $0.878$ & \citet{torres_t1} \\[2pt]
    $R_\star$ & R$_{\odot}$ & $0.807$ & \citet{torres_t1} \\[2pt]
    $T_{\rm eff}$ & K & $5230$ & \citet{torres_t1} \\[2pt]
    $\log(g)$ & cgs & $4.57$ & \citet{torres_t1} \\[2pt]
    $a/R_\star$ & & $10.52$ & \citet{torres_t1} \\[2pt]
    $i$ & degrees & $90.0$ & \citet{torres_t1} \\[2pt]
    $P$ & days & $3.03007008$ & \citet{kokori_exoclockIII}$^{\,*}$ \\[2pt]
    $a$ & AU & $0.03924$ & This work \\[2pt] 
    $M_p$ & M$_\mathrm{J}$ & $0.697$ & \citet{Bonomo2017} \\[2pt]
    $R_p$ & R$_\mathrm{J}$ & $1.067$ & \cite{Bonomo2017} \\[2pt]
    $\mu$ & mas yr$^{-1}$ & $38.1251$ & \href{https://gea.esac.esa.int/archive/}{Gaia Archive (DR3)} \\[2pt]
    $D$ & pc & $159.62$ & \href{https://gea.esac.esa.int/archive/}{Gaia Archive (DR3)} \\[2pt]
    $P_{\rm rot,\star}$ & days & $40.2$ & \citet{Dittmann2009} \\[3.5pt]   
    \hline \hline
    \multicolumn{4}{l}{$^{*}$\footnotesize{Orbital period used in the light curve model fits}} \\[2.5pt] 
\end{tabularx}
\caption{Adopted properties of the TrES-1 system.}
\label{tab:properties}
\end{table}

While the published data of TrES-1\,b are already extensive, the data set can still be improved, particularly in developing a fuller baseline of transit observations. For this, we turn to the diverse and talented community of citizen scientists, who have a rich history of contributing to exoplanet research. We include dozens of citizen scientist light curves from three separate databases in our analysis of TrES-1\,b. The oldest contributions are from the Exoplanet Transit Database (ETD), which was established in 2008 by the Variable Star and Exoplanet Section of the Czech Astronomical Society \citep{poddany_etd_2010,poddany_etd_2011}. Transit observations from this site have been used for ephemeris refinement in numerous publications \citep[e.g.,][]{Baluev2015,Mallonn2019,edwards_orbytsI,Hagey2022}. The remaining citizen science light curves are from two related programs, ExoClock \citep{kokori_exoclockI,kokori_exoclockII, kokori_exoclockIII} and Exoplanet Watch \citep{Zellem2020, Pearson2022}, which collate observations with the specific purpose of ensuring up-to-date ephemerides for planets that are candidates for atmospheric characterization. 

Altogether, we collected -- and in many cases refit -- 123 transit observations of TrES-1\,b, spanning 18 years. We describe the data, including the light curve fitting process, in Section\>\ref{sec:data}. The models and fitting approach are described in Section\>\ref{sec:model_fitting}, and the results are presented in Section\>\ref{sec:results}. In Section\>\ref{sec:interpretation} we address the implications of our results, considering multiple physical interpretations, and our conclusions are given in Section\>\ref{sec:conclusion}.

\section{Data}\label{sec:data}
The data used in this study include $123$ transit midtimes (Appendix\>\ref{appendix:transit_data}), five eclipse midtimes (Appendix\>\ref{appendix:eclipse_data}), and $37$ radial velocity measurements (Appendix\>\ref{appendix:rv_data}). We convert all time stamps to the BJD$_\text{TDB}$ timing standard, which helps to avoid systematic timing errors \citep{Eastman2010}.

\subsection{Transit Light Curve Fitting}\label{sec:lightcurve}
The transit midtimes used in this study are listed in Appendix\>\ref{appendix:transit_data} with the associated source and, when applicable, contributing observers. Most of the transit times in our analysis are from ground-based observations, but we also include data from TESS, the Hubble Space Telescope (HST), and Spitzer. The variety of contributing observatories and studies may introduce inconsistencies that become important for detailed timing analyses. Thus, instead of relying on previously reported midtransit times, we elect to refit data wherever practicable. 

Light curves from several published studies are available through public archives such as Visier\footnote{\href{https://vizier.cds.unistra.fr}{https://vizier.cds.unistra.fr}} or the NASA Exoplanet Archive\footnote{\href{https://exoplanetarchive.ipac.caltech.edu}{https://exoplanetarchive.ipac.caltech.edu}} \citep{Christiansen2025}. For some cases where the light curves are unavailable, we contacted the lead-authors directly, obtaining data from \citet{Narita2007}, \citet{Winn2007}, \citet{Sada2012}, and \citet{Hrudkova2009}. Where the light curves were still unavailable, we used the midtimes directly as indicated in the table in Appendix\>\ref{appendix:transit_data}. We also compiled and refit transit observations from ETD, Exoclock, and Exoplanet Watch up to 2022-Nov-05. Many of these transits were refit and published first in \citet{kokori_exoclockIII} using the same procedure we describe below. Thus, in these cases we cite the midtimes from \citet{kokori_exoclockIII} and indicate the contributing observer in Appendix\>\ref{appendix:transit_data}. The ETD light curves were downloaded from the VarAstro database\footnote{\href{https://var.astro.cz/en/Exoplanets/20}{https://var.astro.cz/en/Exoplanets/20}} \citep{VarAstro}, with the corresponding light curve ID numbers also provided in Appendix~\ref{appendix:transit_data}.

We also use TESS data from Sector 14 (2019-Jul-18 to 2019-Aug-15), which was included in the \citet{kokori_exoclockIII} analysis, Sector 40 (2021-Jun-24 to 2021-Jul-23), Sector 41 (2021-Jul-23 to 2021-Aug-20), Sector 53 (2022-Jun-13 to 2022-Jul-09), and Sector 54 (2022-Jul-09 to 2022-Aug-05). In each of these sectors, data with a two minute cadence are available, and we downloaded the corresponding Pre-search Data Conditioning (PDC) light curves \citep{smith_pdc,stumpe_pdc1,stumpe_pdc2}. Using the ephemerides from \citet{kokori_exoclockIII}, we identify transits in the TESS data by using windows centered around the expected midtransit times, with a window width of 3 times the transit duration to provide sufficient baseline before and after the transit. 

We analyze all light curves using the Python package \texttt{pylightcurve} \citep{pylightcurve}, which performs model fits with an affine invariant Markov chain Monte Carlo (MCMC) ensemble sampler that is implemented by the \texttt{emcee} Python package \citep{emcee}. The free parameters in our light curve fits are the planet-to-star radius ratio ($R_p / R_\star$) and the midtransit time for the given epoch. To account for out-of-transit systematics, we apply a quadratic detrending model. The fixed transit parameters, listed in Table \ref{tab:properties}, are adopted from \citet{torres_t1}, with the ephemeris taken from \citet{kokori_exoclockIII}. We further use the quadratic limb-darkening law from \citet{Claret2013}, implemented via the \texttt{ExoTETHyS} package \citep{exotethys}, with coefficients calculated based on the specific filter that was used for each observation and the relevant stellar parameters in Table\>\ref{tab:properties}.

For each set of observations, we use the \texttt{curve\_fit} function from the \texttt{scipy} Python package \citep{scipy} to perform an initial fit. We then use the root mean square (RMS) of the residuals as the 1-$\sigma$ uncertainty when fitting the corresponding data with the \texttt{pylightcurve} package. Such a scaling procedure is commonly used \citep[e.g., in the analysis of HST data][]{Tsiaras2018}. A light curve is rejected at this stage if the measured midtransit time is not within 20 minutes of the linear ephemeris from \citet{kokori_exoclockIII}, as such large discrepancies are indicative of poor-quality data resulting from various issues like incomplete transit coverage or significant systematics. The 40-minute window around the expected mid-time is sufficiently conservative to retain any plausible timing variations while removing unreliable observations. Furthermore, if the midtransit time or planet-to-star radius ratio do not converge, the light curve is not used in the study. We consider the results to be converged if the magnitude of the positive and negative uncertainties are within 50\% of each other. We note that the ExoClock database already has a first-pass data quality filter, using the same metrics. 

Most data sets required no special treatment, but two did. The first was the light curve from \citet{Narita2007}, because the exposure time varied across the transit light curve. Specifically, the first 139 exposures were taken with a 40 second exposure time, before switching to a 60 second exposure time due to changing observing conditions. In this case, we treat each portion separately in terms of uncertainties, but fit them simultaneously, with a single midtransit time and planet-to-star radius ratio. The resulting midtransit time and uncertainty were consistent with those from \citet{Narita2007} to 1-$\sigma$.

The Hubble Space Telescope (HST) data \citep{rabus_t1_2009_hst} also require particular attention. Three transits were observed in November 2004, January 2005, and March 2005 with the ACS/HRC instrument using the grism G800L (0.5794--1.0251 $\mu$m). Each transit spans five HST orbits, with the third and fourth orbits capturing the transit and the others providing the out-of-transit baseline. The first transit is contaminated by a spot-crossing event that caused a flux increase during the third HST orbit. To mitigate biases in the transit fit we exclude the obviously contaminated data points. We validate this approach by confirming that removing the third orbit entirely has negligible impact on the derived transit mid-time, discussed below.

While individual fits of the HST transits yield mid-time uncertainties of 3–4 seconds, the observed HST transit times deviate significantly (by $\sim1$ minute) from predictions based on the current linear ephemeris \citep{kokori_exoclockIII}. This may be due to data gaps caused by Earth obscuration, which lead to incomplete transit coverage. We also observe variability in the best-fit planet-to-star radius ratio across individual model fits. Thus, we choose to fit all three HST transit observations simultaneously to achieve complete transit coverage. This method provides a single midtransit time, centered on the January 2005 observation.

For all transit fits, two additional criteria led to the final set of 123 transit midtimes listed in Appendix\>\ref{appendix:transit_data}. First, we included only midtimes with uncertainties below approximately 1 minute ($0.0007$ days), as larger uncertainties are often symptomatic of problematic observations that can compromise the mid-time determination due to issues such as incomplete transit coverage, misreported time standards, or other data quality problems that are difficult to trace. Second, we required that the ``coverage'' of each transit observation be at least 90\%, which helps to exclude partial transits and poorly sampled data. To define the transit coverage, we first estimated the times and duration of ingress and egress. We then calculated the expected number of observations during these periods using the median observing cadence, as some data sets have varying cadences. If the number of actual observations was less than 90\% of this expected count, the transit was excluded. The same process was applied to the entire transit duration, taken to be 2.6 hours centered on the best-fit midtransit time.

\subsection{Eclipse Midtimes}\label{sec:eclipse_times}
There are five eclipse observations of TrES-1\,b in the literature, taken with Spitzer IRAC \citep{Charbonneau2005}, for which we adopted the midtimes from the analysis of \citet{Cubillos2014}. Though the eclipse times used in this study are sourced directly from \citet{Cubillos2014}, a light travel time correction of $39.2$ seconds had to be subtracted to account for the time delay across the orbit. This time delay was calculated as $2a/c$, where $a$ is the semi major axis and $c$ is the speed of light, under the assumption of a circular orbit. Given the small eccentricities inferred in our various model fits, more precise corrections were found to be unnecessary. The corrected eclipse midtimes are listed in Appendix\>\ref{appendix:eclipse_data}.

\subsection{Radial Velocity Observations}\label{sec:rv_data}
There are 48 radial velocity (RV) measurements in the literature for TrES-1\,b from four published studies, covering the years 2004--2016. These include seven measurements from \citet{Alonso2004} spanning 18 days, five measurements from \citet{Laughlin2005} spanning three days, 20 measurements from \citet{Narita2007} who focused on measuring the Rossiter-Mclaughlin effect, and 15 measurements from \citet{Bonomo2017} spanning 2.8 years. Because the \citet{Narita2007} study focused on measuring the Rossiter-McLaughlin effect, some of the measurements are skewed during the transit of TrES-1\,b. This was noted by \citet{Cubillos2014} and, following their example, we exclude the corresponding data points. Excluding those points leaves a total of 37 RV measurements, listed in Appendix\>\ref{appendix:rv_data}, with the time standards converted from HJD$_{\rm UTC}$ to BJD$_{\rm TDB}$ where appropriate.

\section{Model Fitting}\label{sec:model_fitting}
All model fits in this study were performed using the \texttt{OrbDot}\footnote{\href{https://github.com/simonehagey/orbdot}{https://github.com/simonehagey/orbdot}} Python package, which is specifically designed to analyze long-term variations in exoplanet orbits \citep{orbdot}. To explore the parameter space, \texttt{OrbDot} employs the nested sampling algorithm \citep{Skilling2006,Feroz2008} via the \texttt{nestle} package \citep{nestle}.\footnote{The \texttt{pymultinest} package \citep{pymultinest,multinest} is also available for nested sampling applications in \texttt{OrbDot}.} For all model fits, we set the number of live points to $1000$ and the evidence tolerance to $0.01$.

In Sections \ref{sec:timing_models} and \ref{sec:rv_models} we describe the transit and eclipse timing and radial velocity models, all of which are implemented in \texttt{OrbDot}. One of the main advantages of \texttt{OrbDot} is its support for joint model fitting, which enables simultaneous fitting of multiple data types to better constrain shared parameters. This technique requires that all models share the same coordinate system, which we define explicitly in Section \ref{sec:coordinate_system}. Finally, we describe the priors chosen for the various model fits in Section\>\ref{sec:priors} and outline our approach to model comparison in Section\>\ref{sec:model_comparison}.

\begin{figure}
    \includegraphics[width=1.0\columnwidth]{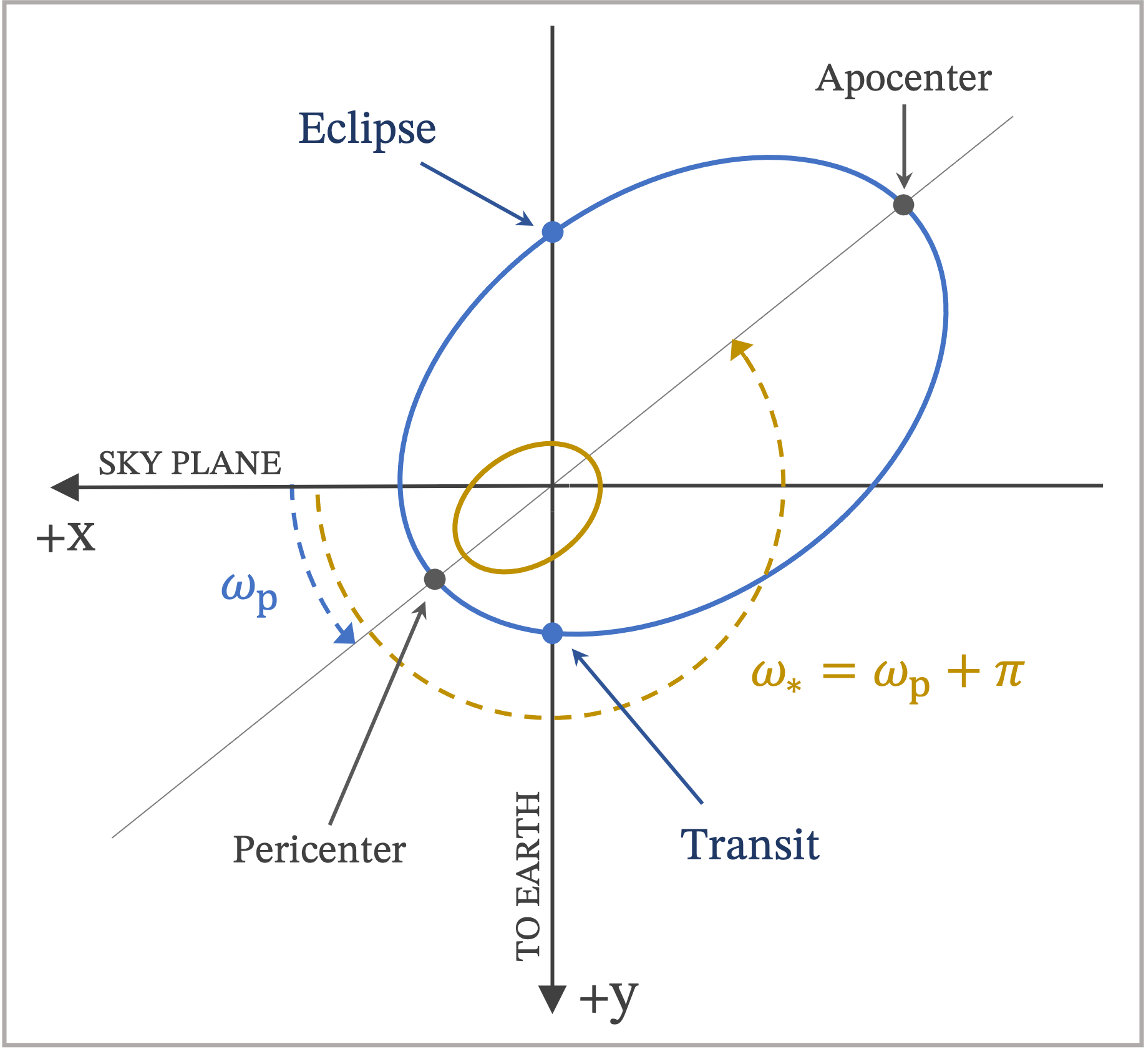}
    \caption{Coordinate system used in the modelling by \texttt{OrbDot}. The positive y--axis points toward the observer, while the x-- and z--axes define the sky plane. The planet's orbit is shown in blue with the argument of pericenter ($\omega_p$) measured from the positive $x$--axis along the direction of motion. The transit and secondary eclipse points are labeled.}
    \label{fig:coordinates}
\end{figure}

\subsection{Coordinate System}\label{sec:coordinate_system}
Figure\>\ref{fig:coordinates} is a diagram of the \texttt{OrbDot} coordinate system, in which the sky plane is defined as the x--z plane, with the y--axis pointing toward the observer along the line of sight. The argument of pericenter of the planet's orbit, denoted here as $\omega_{p}$, is determined from the positive x--axis, such that a transit occurs when the true anomaly $\phi$ is equal to $\phi_{_{\rm I}}=\frac{\pi}{2}-\omega_{p}$ and an eclipse occurs when $\phi_{_{\rm II}}=\frac{3\pi}{2}-\omega_{p}$. 

\subsection{Transit and Eclipse Timing Models}\label{sec:timing_models}
For a planet on a circular orbit with a constant orbital period, we expect the midtimes of transit ($t_{_{\rm I}}$) and eclipse ($t_{_{\rm II}}$) light curves to increase linearly with transit epoch. These \textbf{constant-period} timing equations are
\begin{eqnarray}
\label{eq:linear_tra}
t_{_{\rm I}} &=& t_{_0} + PE, \\ 
t_{_{\rm II}} &=& t_{_0} + PE + \frac{P}{2},
\label{eq:linear_ecl}
\end{eqnarray}
where $t_{_0}$ is a reference transit time and $P$ is the orbital period. The epoch, $E$, represents the number of orbits that have passed since time $t_{_0}$. By fitting this linear model and subtracting it from the data, we can detect deviations from a circular, constant-period model.

The first secular evolution model that we explore is a planet on a circular orbit, but undergoing \textbf{orbital decay}. This leads to a gradual shrinking of the orbit, accompanied by a decreasing orbital period and a corresponding shift in the timing of the transit and eclipse centers. This behavior is modeled by introducing a quadratic term into Equations \ref{eq:linear_tra} and \ref{eq:linear_ecl}:
\begin{eqnarray} \label{eq:decay}
\label{eq:decay_tra}
t_{_{\rm I}} &=& t_{_0} + PE + \frac{1}{2}\frac{dP}{dE}E^2, \\
t_{_{\rm II}} &=& t_{_0} + PE + \frac{P}{2} + \frac{1}{2}\frac{dP}{dE}E^2,
\label{eq:decay_ecl}
\end{eqnarray}
where $dP/dE$ is the rate of change of the period in units of days per epoch. The rate is negative for orbital decay.

In the models presented so far, the orbits remain strictly circular. For eccentric orbits, the term $\frac{2P}{\pi} e\cos{\omega_p}$ must be added to the eclipse midtimes to address timing offsets that arise from variations in the planet's speed at different positions in the orbit.

The second evolutionary model we explore is \textbf{apsidal precession}, which requires the orbit to have at least some eccentricity. In this case the argument of pericenter evolves at a constant rate, inducing long-term sinusoidal trends in the transit and eclipse timing data. At any given transit epoch, the planet's argument of pericenter, $\omega_{p}$, is given by
\begin{equation} \label{eq:omega_p}
\omega_{p}\left(E\right) = \omega_{_0} + \frac{d\omega}{dE}E,
\end{equation}
where $\omega_{_0}$ is the value of $\omega_{p}$ at the reference time $t_{_0}$, and $d\omega/dE$ is the precession rate in units of radians per orbit.

We model apsidal precession following the formalism of \citet{Patra2017}, who derived an approximation of the precession that is first-order in eccentricity using the results of \citet{Gimenez1995}. Specifically: 
\begin{eqnarray} \label{eq:precession_tra}
t_{_{\rm I}} &=& t_{_0} + P_s E - \frac{e P_a}{\pi}\cos{\omega_{p}}, \\
\label{eq:precession_ecl}
t_{_{\rm II}} &=& t_{_0} + P_s E+ \frac{P_a}{2} + \frac{eP_a}{\pi}\cos{\omega_{p}},
\end{eqnarray}
where $P_a$ is the ``anomalistic'' orbital period -- the elapsed time between subsequent pericenter passages, which characterizes the osculating orbit -- and $P_s$ is the sidereal period, which is related to the anomalistic period by
\begin{equation}\label{eq:anomalistic_period}
P_s = P_a\left(1-\frac{d\omega/{dE}}{2\pi}\right).
\end{equation}
The sidereal period represents the ``observed'' orbital period of the system, that is, the elapsed time between subsequent transits.

\subsection{Radial Velocity Models}\label{sec:rv_models}
The observed RV of a star that hosts a massive planet will exhibit periodic variations as the star orbits the center of mass of the system. The amplitude of these variations, denoted $K$, is given by:
\begin{equation} \label{eq:RV_amplitude}
K=\left(\frac{2 \pi G}{P}\right)^{1/3} \frac{M_p \sin i}{\left(M_{\star}+M_p\right)^{2/3}} \frac{1}{\sqrt{1-e^2}},
\end{equation}
where $M_p$ is the planet mass, $M_{\star}$ is the stellar mass, $i$ is the orbital inclination relative to the line of sight, and $G$ is the gravitational constant.

At any given time, the RV signal depends on the position of the planet in its orbit, determined by the true anomaly $\phi$, and the systemic velocity along the line of sight, denoted $\gamma$. To account for potential instrumental offsets, \texttt{OrbDot} fits a separate $\gamma_j$ parameter for each data set (denoted by subscript $j$), which is standard practice when combining RV data from multiple sources. Additionally, we include first- and second-order acceleration terms, $\dot{\gamma}$ and $\ddot{\gamma}$, respectively, to account for the possible influence of a distant, non-resonant companion with an orbital period longer than the observational baseline. 

Thus, the total observed RV signal is modeled as
\begin{equation} \label{eq:RV}
\begin{aligned}
\centering
v_r &=& K[\cos{(\phi\left(t\right)+\omega_p)}+e\cos{\omega_p}] \\ 
&+& \gamma_j + \dot{\gamma} \left(t-t_0\right) + \ddot{\gamma} \left(t-t_0\right)^2.
\end{aligned}
\end{equation}
The use of the planet's argument of pericenter, $\omega_p$, is consistent with the angles defined in our coordinate system (Figure\>\ref{fig:coordinates}), such that a positive RV corresponds to a redshift of the star. 

To avoid underestimating parameter uncertainties, we incorporate an instrument-dependent “jitter” term, $\sigma_j$, which accounts for additional noise from stellar activity or instrumental systematics. This jitter term is added in quadrature with the individual measurement errors in the log-likelihood calculation.

To compute the true anomaly $\phi$ at any observation time requires the time of pericenter passage relative to the most recent transit time ($t_{_{\rm I}}$), as determined by the relevant timing equations from Section \ref{sec:timing_models}. For models involving secular evolution, we adjust the necessary orbital parameters to reflect their value at $t_{_{\rm I}}$. For apsidal precession, we use the precession model’s predicted argument of pericenter (Equation \ref{eq:omega_p}) and anomalistic period (Equation \ref{eq:anomalistic_period}), while for orbital decay, we adjust the period.\footnote{The RV semi-amplitude $K$ (Equation \ref{eq:RV_amplitude}) is also, in principle, time-dependent under orbital decay. However, the variation is negligible over the timescales of the available data and is therefore not included in this analysis.} This allows us to consistently update the RV model for long-term dynamical effects. For more detail, we refer the reader to the \texttt{OrbDot} \href{https://orbdot.readthedocs.io/}{documentation}.

\subsection{Priors}\label{sec:priors}
\begin{figure*}[ht]
    \includegraphics[width=\textwidth]{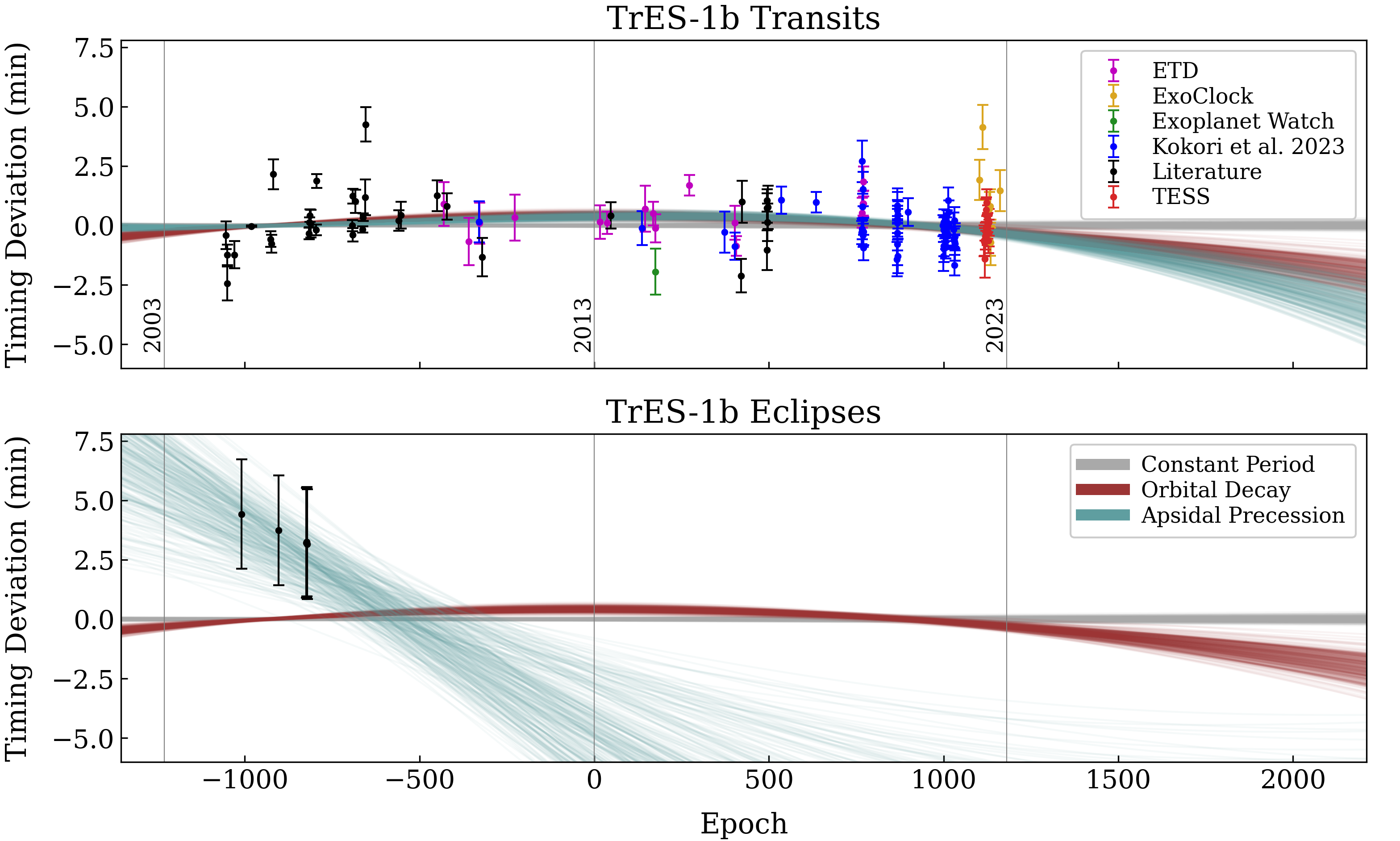}
    \caption{Transit (top) and eclipse (bottom) timing variations for TrES-1 b. Each data point represents the difference between an observed time and the expected midpoint from the best-fit constant-period, circular orbit model. The plot shows 300 random samples from the weighted posterior distributions of the joint model fits -- which included RV data -- for the constant-period, orbital decay, and apsidal precession models.}
    \label{fig:o-c}
    \vspace{0.4cm}
\end{figure*}

The prior distributions for all model parameters are provided in the tables in Section\>\ref{sec:results} and in Appendix \ref{appendix:results_table}. For all models, Gaussian priors were applied to $t_0$ and $P$, based on the known transit ephemeris. Wide uniform priors were used for the orbital decay rate $dP/dE$ and the apsidal precession rate $d\omega/dE$, with the orbital decay rate allowed to be positive (indicating an increasing orbital period) as well as negative. In the RV models, a uniform prior was used for the amplitude $K$, and Gaussian priors were assigned to the systemic velocities $\gamma_j$ based on published analyses. Since the \citet{Alonso2004}, \citet{Laughlin2005} and \citet{Narita2007} data were already zeroed, these priors were centered at zero. A log-uniform prior was applied to the jitter terms $\sigma_j$. Priors on $e$ and $\omega_p$ varied by model. In the constant-period and orbital decay model fits with nonzero eccentricity, $\omega_p$ was fixed at zero to constrain the largest possible $e$. For the apsidal precession model, midtimes indicate that at $t_0$, the timing deviation must be negative (see Figure\>\ref{fig:o-c}), which restricts $\omega_0$ to be in the range $(\frac{\pi}{2},\frac{3\pi}{2})$.

\subsection{Model Comparison}\label{sec:model_comparison}
We use the Bayesian evidence, denoted as $\log{\mathrm{Z}}$, as a fundamental metric for comparing the outcomes of various model fits. A lower $\left|\log{\mathrm{Z}}\right|$ value signifies a better fit to the observed data. The Bayesian evidence is computed directly by the nested sampling algorithm in \texttt{OrbDot}, while the Bayes factor between two models is
\begin{equation}\label{eq:bayes_factor}
\log{B_{12}} = \log{\mathrm{Z}}_{1} - \log{\mathrm{Z}}_{2}.
\end{equation}
We determine the preferred model by comparing the Bayes factor to the thresholds established by \citet{kass_and_raftery}, described in Table\>\ref{tab:bayesian_evidence}.

\begin{table}[ht]
    \addtocounter{table}{-1}
    \centering
    \begin{tabularx}{0.48\textwidth}{ll}
        \multicolumn{2}{l}{} \\
        \normalsize{\textbf{Condition}} & \normalsize{\textbf{Evidence for Model 1 ($M_1$)}}  \\ \noalign{\vskip 0.5mm}
        \hline\hline \noalign{\vskip 1mm}
        $B_{12} \leq 1$ &  $M_1$ is not supported over $M_2$  \\ \noalign{\vskip 1mm}
        $1 < B_{12} \leq 3$ &  Evidence for $M_1$ is barely worth mentioning \\ \noalign{\vskip 1mm}
        $3 < B_{12} \leq 20$ &  Positive evidence for $M_1$ \\ \noalign{\vskip 1mm}
        $20 < B_{12} \leq 150$ &  Strong evidence for $M_1$ \\ \noalign{\vskip 1mm}
        $150 < B_{12}$ &  Very strong evidence for $M_1$ \\ \noalign{\vskip 1mm}
        \hline\hline
    \end{tabularx}
    \caption{Bayesian evidence thresholds from \citet{kass_and_raftery} for comparing Model 1 ($M_1$) vs. Model 2 ($M_2$).}
    \label{tab:bayesian_evidence}
\end{table}

\begin{table}
  \addtocounter{table}{-1}
\centering
    \begin{tabularx}{\linewidth}{l|ccc}
             \multicolumn{1}{l}{} & \multicolumn{3}{c}{\normalsize{\textbf{Bayesian Evidence}}} \\
             \multicolumn{1}{l}{\normalsize{\textbf{Model}}} & \multicolumn{1}{c}{\textbf{All Data}} & \multicolumn{1}{c}{\textbf{No Eclipse}} & \multicolumn{1}{c}{\textbf{Mid-Times}} \\[1pt] \hline\hline
        \rule{0pt}{2.7ex}Apsidal precession & $-334.6$ & $-329.8$ & $-199.7$ \\[3pt] 
        Orbital decay, eccentric &  $-336.4$ & -- & $-200.9$ \\[3pt]
        Orbital decay, circular & $-340.2$ & $-334.0$ & $-204.6$ \\[3pt]
        Constant period, eccentric & $-342.2$ & -- & $-206.6$ \\[3pt]
        Constant period, circular & $-345.4$ & $-339.5$ & $-210.2$ \\[2.7pt] \hline\hline
    \end{tabularx}
\caption{Comparison of orbital models for TrES-1 b, including constant period (circular and eccentric orbits), orbital decay (circular and eccentric), and apsidal precession. Each model was fit to different data subsets: (1) all available data (transit, eclipse, and RV), (2) data without eclipses (transit and RV only), and (3) timing data (transit and eclipse only). The full results from all model fits are in Appendix\>\ref{appendix:results_table}. Within each column, models are ranked from best to worst fit based on Bayesian evidence  ($\log{\mathrm{Z}}$), with apsidal precession identified as the best model in all cases. Note that Bayesian evidence values should only be compared within each column, not across different data subsets.}
\label{tab:results_summary}
\end{table}

\section{Results}\label{sec:results}
Altogether, we fit five models to the TrES-1\,b data:

\begin{enumerate}[(i)]
    \item a constant-period, circular orbit.\vspace{-5pt}
    \item a constant-period, slightly eccentric orbit.\vspace{-5pt}
    \item a decaying, circular orbit.\vspace{-5pt}
    \item a decaying, slightly eccentric orbit.\vspace{-5pt}
    \item a slightly eccentric, precessing orbit.
\end{enumerate}

We included ``slightly eccentric'' ($e \ll 0.1$) models to account for the significant offsets observed in the eclipse timing relative to predictions from circular orbit models (see Figure\>\ref{fig:o-c}). Figure\>\ref{fig:o-c} illustrates the transit and eclipse timing deviations compared to the constant-period, circular orbit model, while Figure\>\ref{fig:oc_eccentric} (Appendix\>\ref{appendix:oc_eccentric}) shows deviations relative to the constant-period, eccentric model. We note that the scatter in the nominal eclipse timing variations is small relative to the associated uncertainties, and thus any interpretation of curvature based solely on the eclipse data should be treated with caution.

Though we treat models (iv) and (v) as distinct, it is important to note that a nonzero eccentricity will cause the orbit to precess, regardless of whether or not it is shrinking due to tidal interactions. Fitting a slightly eccentric, decaying orbit model (iv) simply provides a useful comparison to the apsidal precession model (v) while avoiding the complexity of a combined scenario.

The statistical evidence strongly favors the apsidal precession model as the best fit overall, with the eccentric orbital decay model as the closest contender. Table\>\ref{tab:results_summary} summarizes the Bayesian evidence ($\log{\mathrm{Z}}$) values for all five models, with full results provided in Appendix\>\ref{appendix:results_table}. Table\>\ref{table:best_results} presents the best-fit apsidal precession solution, based on a joint fit to all data, yielding an eccentricity of $e=0.0059$, an argument of pericenter $\omega_{p} = 1.909$ rad, and a precession rate of $d\omega/dE = 0.00057$ rad\,/\,E (equal to $4.0\>^{\circ}/{\rm yr}$), which corresponds to a precession period of 90 years.

For the joint model fits (see the `All Data' column in Table\>\ref{tab:results_summary}), evidence for precession is ``very strong'' compared to the circular constant-period model ($B = 4.96 \times 10^4$), the eccentric constant-period model ($B = 1.58 \times 10^3$), and the circular orbital decay model ($B = 2.62 \times 10^2$). Against the eccentric decay model, the evidence is ``positive'' ($B = 5.19$). 

It is clear that allowing for nonzero eccentricity improves the fits for both the constant-period and orbital decay models, with best-fit eccentricity values of $e=0.00128$ and $e=0.00126$, respectively. The best-fit orbital decay rate is $dP/dt = -7.2$ ${\rm ms\,yr^{-1}}$ for the circular orbit model, and $-7.1$ ${\rm ms\,yr^{-1}}$ for the eccentric model. Thus, while including eccentricity in the decay model does not alter the derived period derivative, it does improve the overall model fit. Notably, the evidence for the \textit{circular}, decaying orbit is ``very strong'' compared to the \textit{eccentric} constant-period model ($B=7.38$), reinforcing the detection of secular evolution in general.

Assuming the measured apsidal precession is real, the best-fit decay rate is consistent with expectations. Since precession periods often exceed observational baselines, only a curved segment of the precession is observable, which may also be well described by a quadratic timing model (such as the orbital decay model). For a given apsidal precession solution, the corresponding time derivative of the \textit{observed} orbital period (the time between subsequent transits) is \citep{Rafikov2009}
\begin{equation} \label{eq:pdot_from_w}
    \frac{dP}{dt}_{\rm (apparent)} = \frac{4 \pi\left(\dot{\omega}\right)^2 e \cos \omega_p}{n^2} \frac{(1-e^2)^{3 / 2}}{(1+e \sin \omega_p)^3} \,.
\end{equation}

Evaluating Equation\>\ref{eq:pdot_from_w} with the best-fit precession model (Table\>\ref{table:best_results}) yields a quadratic period variation of $-6.4$ ${\rm ms\,yr^{-1}}$, which is consistent with the best-fit orbital decay model ($-7.1$ ${\rm ms\,yr^{-1}}$, Appendix\>\ref{appendix:results_table}). These findings suggest that the apsidal precession model captures the curvature in the timing data as effectively as the quadratic decay model. Altogether, these results strongly indicate a secular change in the orbit of TrES-1\,b, with apsidal precession as the favored mechanism. The physical implications are discussed further in Section\>\ref{sec:interpretation}.

\begin{table}
  \addtocounter{table}{-1}
    \begin{tabularx}{1.072\linewidth}{|lcc||c|}
        \multicolumn{4}{c}{\normalsize{\textbf{Best TrES-1\>b Solution}}}\\[1.7pt]
        \hline
        & \textbf{Unit} & \textbf{Prior} & \textbf{Credible Interval}  \\ \hline \noalign{\vskip 1.3mm}  \hline
        \rule{0pt}{2.8ex}$t_0$ & BJD$_{_\mathrm{TDB}}$ & $\mathcal{N}(2456368.3807,0.01)$ & $2456368.37915\,^{+\,0.00072}_{-0.00065}$ \\[5pt]
        $P_s$ & days & $\mathcal{N}(3.03007,0.001)$ & $3.03006664\,^{+\,0.00000099}_{-0.00000083}$ \\[5pt]
        $e$ & & $\mathcal{U}(0.0,0.01)$ & $0.0059\,^{+\,0.0026}_{-0.0028}$ \\[5pt]
        $\omega_0$ & rad & $\mathcal{U}(\frac{\pi}{2},\frac{3\pi}{2})$ & $1.909\,^{+\,0.081}_{-0.052}$ \\[5pt]
        $\dot{\omega}$ & rad\,/\,E & $\mathcal{U}(0.0,0.001)$ & $0.00058\,^{+\,0.00016}_{-0.00009}$ \\[5pt]
        $\dot{\omega}$ & $^\circ$\,/\,yr & Derived & $4.0\,^{+\,1.1}_{-0.6}$ \\[5pt]
        $K$ & m\,/\,s & $\mathcal{U}(100,120)$ & $109.2\,^{+\,1.8}_{-1.8}$ \\[5pt]
        $\gamma_A$ & m\,/\,s & $\mathcal{U}(-10,10)$ & $-2.1\,^{+\,3.9}_{-3.8}$ \\[5pt]
        $\gamma_B$ & m\,/\,s & $\mathcal{U}(-20470,-20440)$ & $-20458.6\,^{+\,1.8}_{-1.8}$ \\[5pt]
        $\gamma_L$ & m\,/\,s & $\mathcal{U}(-10,10)$ & $-5.4\,^{+\,1.5}_{-1.5}$ \\[5pt]
        $\gamma_N$ & m\,/\,s & $\mathcal{U}(-10,10)$ & $-0.4\,^{+\,3.6}_{-3.5}$ \\[4pt]
        $\sigma_A$ & m\,/\,s & Fixed & $0.34$ \\[1.7pt]
        $\sigma_B$ & m\,/\,s & Fixed & $6.3$ \\[1.7pt]
        $\sigma_L$ & m\,/\,s & Fixed & $0.39$ \\[1.7pt]
        $\sigma_N$ & m\,/\,s & Fixed & $0.24$ \\[2.1pt]
        \hline\hline
    \end{tabularx}
\caption{Joint fit results for the apsidal precession model, identified as the preferred model for TrES-1\,b, including 68\% credible intervals for each parameter. The nomenclature for the prior distributions is as follows: $\mathcal{U}$ represents a uniform prior, defined by the upper and lower bounds, and $\mathcal{N}$ is a Gaussian (normal) prior, defined by the mean and standard deviation. For the instrument-dependent parameters $\gamma$ and $\sigma$, subscripts indicate the data source (i.e., \citealt{Alonso2004}, \citealt{Bonomo2017}, \citealt{Laughlin2005}, and \citealt{Narita2007}). The jitter values ($\sigma_j$) were fixed based on a prior circular RV model fit that was conducted before the joint fit.}
\label{table:best_results} 
\end{table}


\subsection{Exploring Different Data Subsets} \label{sec:results_subsets}
We tested each model on different data subsets to better understand whether the solutions are driven by particular data types. Those subsets are (1) all data (transit, eclipse, and RV), (2) data without eclipses (transit and RV), and (3) timing data only (transit and eclipse). Table\>\ref{tab:results_summary} summarizes the Bayesian evidence ($\log{\mathrm{Z}}$) for the models, with each column ranked by evidence for that subset of data. Note that the $\log{\mathrm{Z}}$ values should only be compared within columns, not across subsets. In all subsets the natural ranking order of the models by evidence is the same, and the apsidal precession model consistently emerges as the best model to describe the TrES-1\,b data.

\subsubsection{Timing Data Only}
When fitting only the transit and eclipse midtimes, the credible intervals for the model parameters remain consistent with those from the joint fits (see Appendix\>\ref{appendix:results_table}). We again find that the evidence for apsidal precession against both of the constant-period models (circular and eccentric), as well as the circular orbital decay model, is ``very strong.'' However, in this case the evidence for the apsidal precession model against the eccentric orbital decay model is ``barely worth mentioning'' ($B = 2.46$). This shows that including RV data in the analysis strengthens the evidence for precession, which highlights the importance of combining multiple data types.

We also tested the impact of excluding three transit midtimes with the smallest uncertainties by far (epochs $-981$, $-663$, and $-662$; see Appendix \ref{appendix:transit_data}), including the mid-time derived from the combined fit of three HST transits (Section \ref{sec:lightcurve}). Our findings remain unchanged, reinforcing the robustness of our model fitting to the TrES-1\,b transit midtimes.

\subsubsection{Excluding Eclipse Times}
Without the eclipse midtimes, the eccentric orbit models (both constant-period and orbital decay) are not well-constrained, as the inferred eccentricity is too small to be significant in the RV measurements. Thus, we do not assess those models when the eclipse midtimes are excluded. However, we still fit the apsidal precession model, as we expect that the changing angular orientation may have an effect on the long-term evolution in RV. When eclipse midtimes are excluded, evidence for precession remains ``very strong'' against the constant-period model ($B = 1.66 \times 10^4$) and ``strong'' against the orbital decay model ($B = 6.60 \times 10^1$).

The orbital decay fit without eclipse midtimes yields $dP/dt = -7.1\,^{+\,1.5}_{-1.5}$ ${\rm ms\,yr^{-1}}$, which is the same as the joint fit results (Appendix\>\ref{appendix:results_table}). The best-fit precession model shows slight differences, with a phase shift ($\omega_0 = 3.33\,^{+\,0.69}_{-0.82}$ rad) and a lower nominal eccentricity that is poorly constrained ($e = 0.0035\,^{+\,0.0032}_{-0.0019}$). Despite these differences, the precession rate ($d\omega/dE = 0.00052\,^{+\,0.00024}_{-0.00015}$ rad/E) remains consistent with the joint fit. 

These findings indicate that including the eclipse midtimes improves the parameter constraints, but they may not be essential for detecting apsidal precession if you have a long baseline of transit and RV observations.

\begin{table}
\addtocounter{table}{-1}
\begin{tabularx}{1.04\linewidth}{|lcc||c|}
    \multicolumn{4}{c}{\normalsize{\textbf{Radial Velocity Results}}}\\[1.7pt]
    \hline
    & \textbf{Unit} & \textbf{Prior} & \textbf{Credible Interval}  \\ \hline \noalign{\vskip 1.5mm} 
    
    \multicolumn{4}{l}{\small{\textbf{TrES-1 b -- Circular Orbit}}}  \\  \hline
    \rule{0pt}{2.6ex}$t_0$ & BJD$_{_\mathrm{TDB}}$ & $\mathcal{N}(2456368.3807,0.01)$ & $2456368.3764\,^{+\,0.0069}_{-0.0066}$ \\[5pt]
    $P$ & days & $\mathcal{N}(3.03007,0.001)$ & $3.0300614\,^{+\,0.0000098}_{-0.0000092}$ \\[5pt]
    K & m\,/\,s & $\mathcal{U}(100,120)$ & $109.3\,^{+\,2.0}_{-2.1}$ \\[5pt]
    $\gamma_A$ & m\,/\,s & $\mathcal{U}(-10,10)$ & $-2.0\,^{+\,4.5}_{-4.0}$ \\[5pt]
    $\gamma_B$ & m\,/\,s & $\mathcal{U}(-20470,-20440)$ & $-20458.9\,^{+\,1.9}_{-1.7}$ \\[5pt]
    $\gamma_L$ & m\,/\,s & $\mathcal{U}(-10,10)$ & $-5.3\,^{+\,1.7}_{-1.6}$ \\[5pt]
    $\gamma_N$ & m\,/\,s & $\mathcal{U}(-10,10)$ & $-0.7\,^{+\,4.1}_{-4.2}$ \\[5pt]
    $\sigma_A$ & m\,/\,s & $\mathcal{J}(10^{-1},10^2)$ & $0.3\,^{+\,2.6}_{-0.3}$ \\[5pt]
    $\sigma_B$ & m\,/\,s & $\mathcal{J}(10^{-1},10^2)$ & $6.3\,^{+\,1.8}_{-1.6}$ \\[5pt]
    $\sigma_L$ & m\,/\,s & $\mathcal{J}(10^{-1},10^2)$ & $0.4\,^{+\,2.5}_{-0.4}$ \\[5pt]
    $\sigma_N$ & m\,/\,s & $\mathcal{J}(10^{-1},10^2)$ & $0.3\,^{+\,2.0}_{-0.3}$ \\[2.6pt]
    \hline \noalign{\vskip 1.5mm} 
    
    \multicolumn{4}{l}{\small{\textbf{TrES-1 c -- Initial Fit}}}  \\ \hline
    \rule{0pt}{2.6ex}$t_0$ & BJD$_{_\mathrm{TDB}}$ & $\mathcal{U}(2456000,2457300)$ & $2456575\,^{+\,66}_{-74}$ \\[5pt]
    $P$ & days & $\mathcal{U}(1100,1300)$ & $1202\,^{+\,43}_{-23}$ \\[5pt]
    K & m\,/\,s & $\mathcal{U}(0,30)$ & $9.9\,^{+\,2.5}_{-2.1}$ \\[5pt]
    $e$ & & $\mathcal{U}(0,1)$ & $0.66\,^{+\,0.15}_{-0.18}$  \\[5pt]
    $\sigma_B$ & m\,/\,s & $\mathcal{J}(10^{-1},10^2)$ & $0.6\,^{+\,1.4}_{-0.4}$ \\[5pt]
    $\sigma_L$ & m\,/\,s & $\mathcal{J}(10^{-1},10^2)$ & $0.8\,^{+\,2.1}_{-0.6}$ \\[2.6pt]
    \hline \noalign{\vskip 1.5mm} 
    
    \multicolumn{4}{l}{\small{\textbf{TrES-1 c -- Informed Fit}}}  \\ \hline
    \rule{0pt}{2.6ex}$t_0$ & BJD$_{_\mathrm{TDB}}$ & $\mathcal{N}(2456575,74)$ & $2456575\,^{+\,42}_{-43}$ \\[5pt]
    $P$ & days & $\mathcal{N}(1202,43)$ & $1200\,^{+\,26}_{-20}$ \\[5pt]
    K & m\,/\,s & $\mathcal{N}(9.9, 2.5)$ & $10.0\,^{+\,1.6}_{-1.5}$ \\[5pt]
    $e$ & & $\mathcal{N}(0.66,0.18)$ & $0.65\,^{+\,0.10}_{-0.10}$ \\[5pt]
    $\sigma_B$ & m\,/\,s & $\mathcal{N}(0.6,1.4)$ & $0.6\,^{+\,1.2}_{-1.3}$ \\[5pt]
    $\sigma_L$ & m\,/\,s & $\mathcal{N}(0.8,2.1)$ & $0.7\,^{+\,1.8}_{-2.1}$ \\[2.6pt] \hline\hline
\end{tabularx}
\caption{Best-fit radial velocity model parameters for TrES-1\,b and the proposed planetary companion, TrES-1\,c, with 68\% credible intervals for upper and lower limits. For TrES-1\,c, two sets of results are shown: an initial model fit using uniform prior distributions ($\mathcal{U}$), informed by Figures \ref{fig:rv_planet_b} and \ref{fig:periodogram}, and a second fit using normal priors ($\mathcal{N}$), informed by the initial fit. The symbol $\mathcal{J}$ denotes a log-uniform (Jeffreys) prior. For the instrument-dependent parameters $\gamma$ and $\sigma$, subscripts indicate the data source (i.e., \citealt{Alonso2004}, \citealt{Bonomo2017}, \citealt{Laughlin2005}, and \citealt{Narita2007}).}
\label{table:rv_results}
\end{table}

\begin{figure*}
\centering
    \hspace*{-0.7cm}\includegraphics[width=1.1\textwidth]{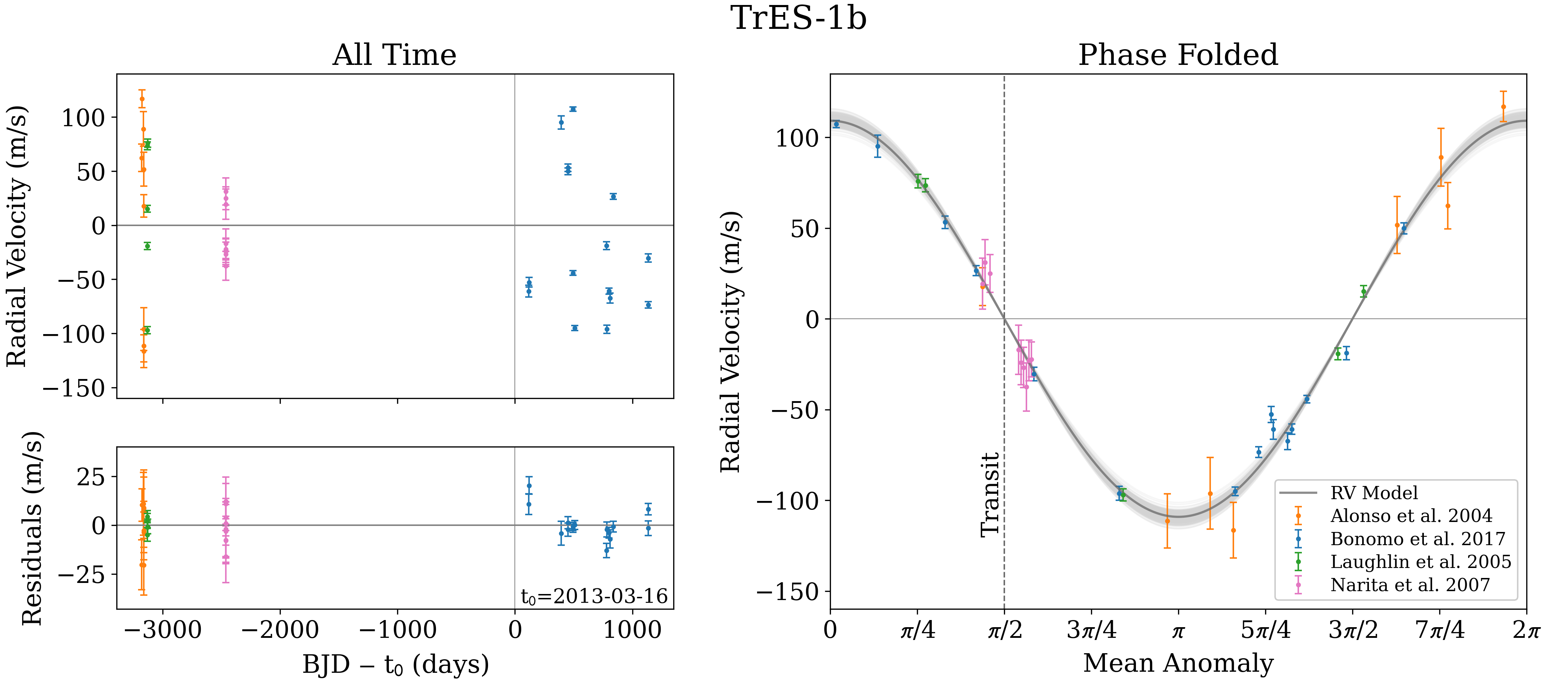}
    \caption{Radial velocity (RV) data and best-fit orbital model for TrES-1\,b. \textbf{Top left:} RV measurements over time, with data points shifted by the systemic RV and observation times shifted to the reference time, $t_0$. \textbf{Bottom left:} Residuals obtained by subtracting the best-fit RV model from the data. \textbf{Right:} Phase-folded RV measurements plotted against mean anomaly, with 300 random samples from the model posterior shown in gray to represent model uncertainty. Data sources are colour-coded as indicated in the legend.}
    \label{fig:rv_planet_b}
\end{figure*}

\begin{figure*}
\centering
    \hspace*{-0.7cm}\includegraphics[width=1.1\textwidth]{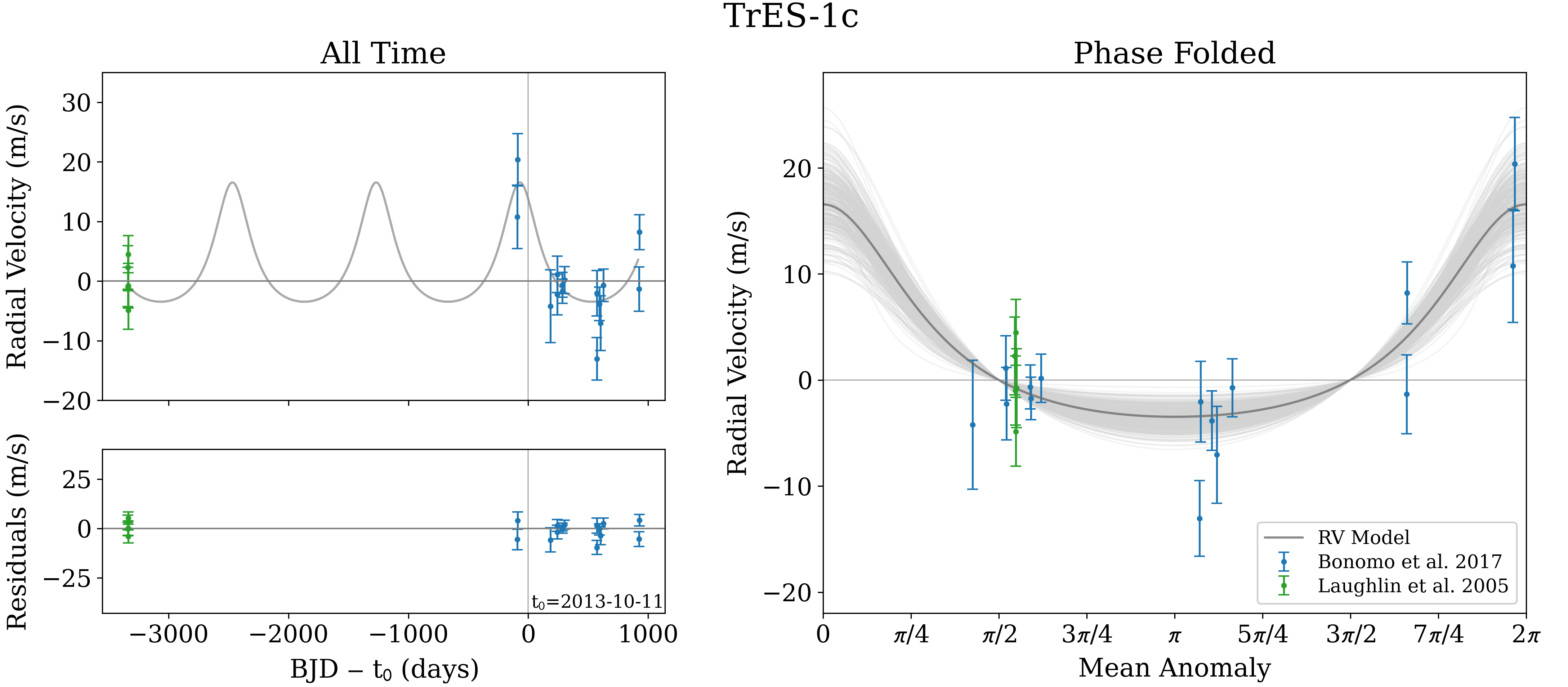}
    \caption{Radial velocity (RV) residuals (see Figure\>\ref{fig:rv_planet_b}) and best-fit orbital model for the proposed planetary companion TrES-1\,c. \textbf{Top left:} RV measurements over time, with data points shifted by the systemic RV and observation times shifted to the reference time, $t_0$. \textbf{Bottom left:} Residuals obtained by subtracting the best-fit RV model from the data. \textbf{Right:} Phase-folded RV measurements plotted against mean anomaly, with 300 random samples from the model posterior shown in gray to represent model uncertainty. Data sources are colour-coded as indicated in the legend.}
    \label{fig:rv_planet_c}
\end{figure*}

\subsection{Radial Velocity Analysis}\label{sec:results_rv}
In addition to the model fits described above, we performed a supplementary analysis of the radial velocity (RV) data alone, partly to search for evidence of additional planets in the TrES-1 system. For this, we fit the following models to the RV data:
\begin{enumerate}[(i)]
	\item a circular orbit.\vspace{-5pt}
	\item a circular orbit with a linear trend.\vspace{-5pt}
    \item a circular orbit with a quadratic trend.\vspace{-5pt}
	\item an eccentric orbit.\vspace{-5pt}
	\item an eccentric orbit with a linear trend.\vspace{-5pt}
    \item an eccentric orbit with a quadratic trend.
\end{enumerate}

The model that best describes the TrES-1\,b signal is the circular orbit model without a long-term trend (i). We find that the Bayesian evidence for this model against all others is ``very strong,'' and that the eccentric model fits are consistent with circular orbits. Although the timing data suggest a small but nonzero eccentricity, the RV measurements are consistent with a circular orbit and lack the precision and sampling required to meaningfully constrain such low eccentricities. The results of this fit, including priors and 68\% credible intervals for the model parameters, are summarized in Table\>\ref{table:rv_results}. Figure \ref{fig:rv_planet_b} shows the RV data with 300 random samples from the weighted posterior distributions.

\begin{figure}
    \hspace*{-0.2cm}\includegraphics[width=1.05\columnwidth]{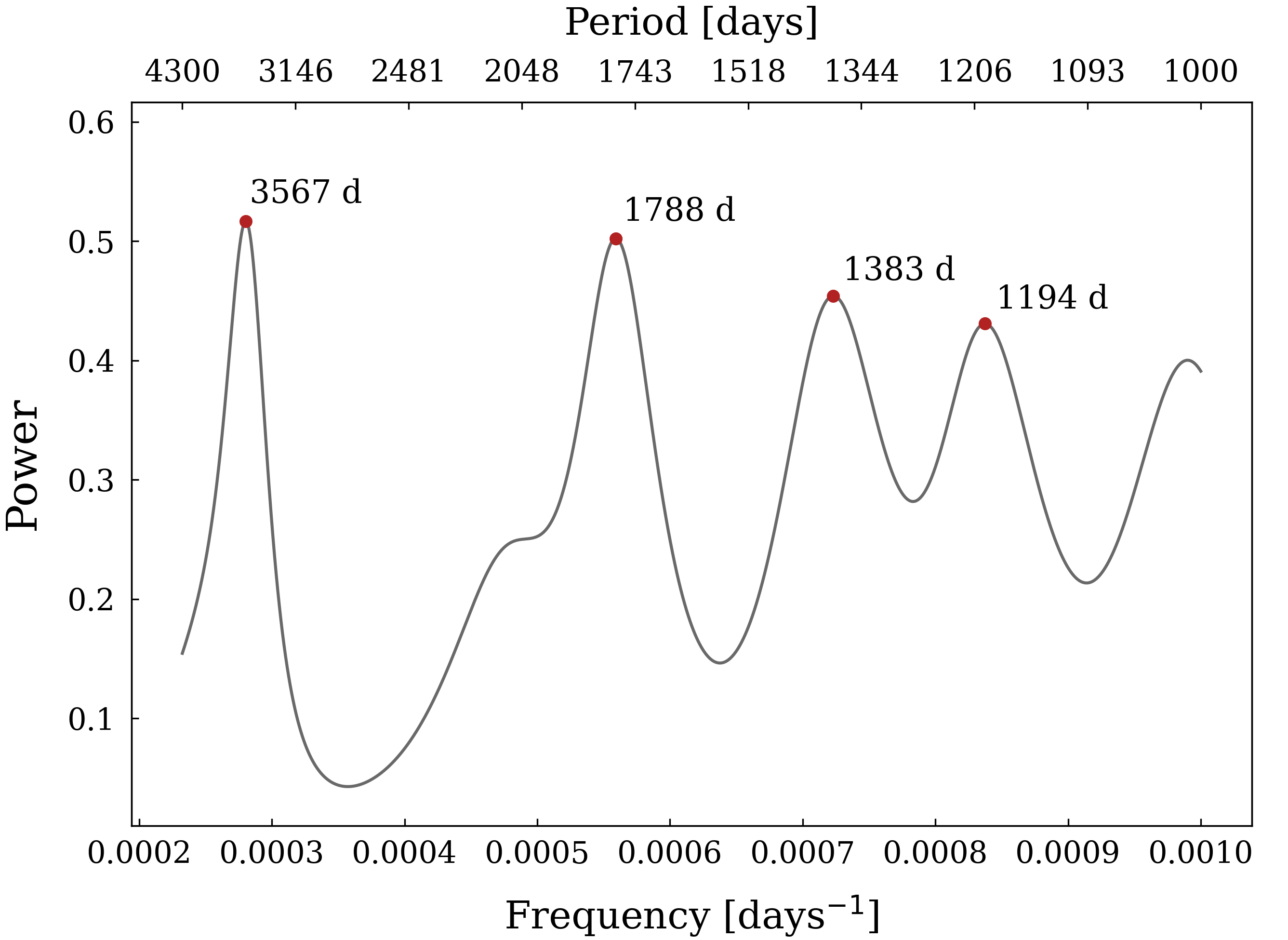}
    \caption{Lomb-Scargle periodogram of the residuals from the TrES-1\,b radial velocity model fit (see Figure \ref{fig:rv_planet_b}). The x--axis displays frequencies in days$^{-1}$ and the corresponding periods are shown on the top axis. The frequency range was selected by restricting the periods to between 1000 and 4300 days, the latter corresponding to the duration of the RV observation baseline. Labeled peaks indicate potentially significant periodic signals with periods of 3567, 1788, 1383, and 1194 days, suggesting possible planetary companions. These peaks informed subsequent radial velocity fits to investigate evidence for a second planet in the TrES-1 system.}
    \label{fig:periodogram}
\end{figure}

The residuals in Figure\>\ref{fig:rv_planet_b} suggest the presence of a possible additional periodic signal. A Lomb-Scargle periodogram of the residuals (Figure\>\ref{fig:periodogram}) identifies peaks at 3567, 1788, 1383, and 1194 days. To explore the possibility of a second planet (TrES-1\,c), we use these peaks to guide initial exploratory fits to the residuals. To this end, we include only the RV data from \citet{Laughlin2005} and \citet{Bonomo2017}, as the measurements from \citet{Alonso2004} and \citet{Narita2007} have uncertainties that are much greater than the residuals (Figure\>\ref{fig:rv_planet_b}). The same models described above (i--vi) were applied to this subset of data.

The best-fitting model for the additional planetary signal is an eccentric orbit with $e=0.65\,^{+\,0.1}_{-0.1}$ and an orbital period of $1200\,^{+\,26}_{-20}$ days. This solution was identified by first performing a model fit with broad uniform priors spanning the periodogram's $1194$--day peak (see Table\>\ref{table:rv_results}), followed by second fit to refine these results by applying Gaussian priors that are centered on the best-fit model parameters from the initial fit, and with the standard deviations matching the largest corresponding uncertainty. The results of this companion planet fit are presented in Table\>\ref{table:rv_results} and plotted in Figure\>\ref{fig:rv_planet_c}. 

In summary, the final TrES-1\,c model has an orbital period of $1200$ days ($2.1$ au), an RV semi-amplitude of $10.0$ ${\rm m\,s^{-1}}$, and an eccentricity of $0.65$. This corresponds to a minimum mass of $M_p \sin i_c \approx 120$ Earth masses ($0.36\,{\rm M}_{\mathrm{J}}$). Further analysis of the residuals after fitting the second planet does not reveal any additional periodic signals.

\section{Interpretation}\label{sec:interpretation}
Our analysis indicates that the preferred orbital model of TrES-1\,b is a slightly eccentric, precessing orbit ($e = 0.0059$, $\omega_p = 1.909$ rad, $d\omega/dE = 0.00058$ rad/E). In this section, we compare the observed precession rate to theoretical expectations (Sections\>\ref{sec:apsidal_precession} and \ref{sec:planet_precession}) and evaluate the implications of other mechanisms, including orbital decay (Section\>\ref{sec:orbital_decay}), potential stellar or planetary mass companions (Sections \ref{sec:line_of_sight} and \ref{sec:outer_planet}), systemic proper motion (Section\>\ref{sec:proper_motion}), and stellar magnetic activity (Section\>\ref{sec:magnetic_activity}).

\subsection{Apsidal Precession}\label{sec:apsidal_precession}
Apsidal precession can arise from several sources, including general relativistic effects and perturbations caused by tidal and rotational distortions. In our assessment of the precession rate, we include the following terms:
\begin{equation*}
    \dot{\omega}_{\rm total} = \dot{\omega}_{\rm GR} + \dot{\omega}_{\rm rot,p} + \dot{\omega}_{\rm rot,\star} + \dot{\omega}_{\rm tide,p} + \dot{\omega}_{\rm tide,\star}.
\end{equation*}
We discuss each of these terms in detail below. However, using the expected values of the preferred model for TrES-1\,b, we find
\begin{equation*}
    \dot{\omega}_{\rm total} = 1.2 \times 10^{-5} \,{\rm rad}~{\rm E}^{-1} = 0.08 \,^\circ\,{\rm yr}^{-1},
\end{equation*}
which is an order of magnitude smaller than the observed precession rate of $4.0^\circ\,{\rm yr}^{-1}$. This suggests that the above standard mechanisms in $\dot{\omega}_{\rm total}$ cannot account for the variations in TrES-1\,b's orbit alone. 

\subsubsection{General Relativity}\label{sec:general_relativity}
The general relativistic contribution to apsidal precession, to the lowest order in eccentricity, is given by
\begin{equation}
    \dot{\omega}_{\rm GR} = \frac{3GM_\star\,n}{ac^2(1-e^2)},
    \label{eq:gr}
\end{equation}
where $G$ is the gravitational constant, $M_\star$ is the mass of the host star, $a$ is the semi-major axis of the planet's orbit, $c$ is the speed of light in a vacuum, $e$ is the eccentricity, and $n = 2\pi/P$ is the mean motion. A detailed derivation of Equation\>\ref{eq:gr} can be found in \citet{Misner1973}. Using the TrES-1 system parameters (Table\>\ref{tab:properties}), we estimate
\begin{equation*}
    \dot{\omega}_{\rm GR} \approx 3 \times 10^{-2} \,^\circ\,{\rm yr}^{-1}.
\end{equation*}

\begin{figure}
    \begin{minipage}{1.0\columnwidth}
        \hspace*{-0.6cm}\includegraphics[width=1.07\textwidth]{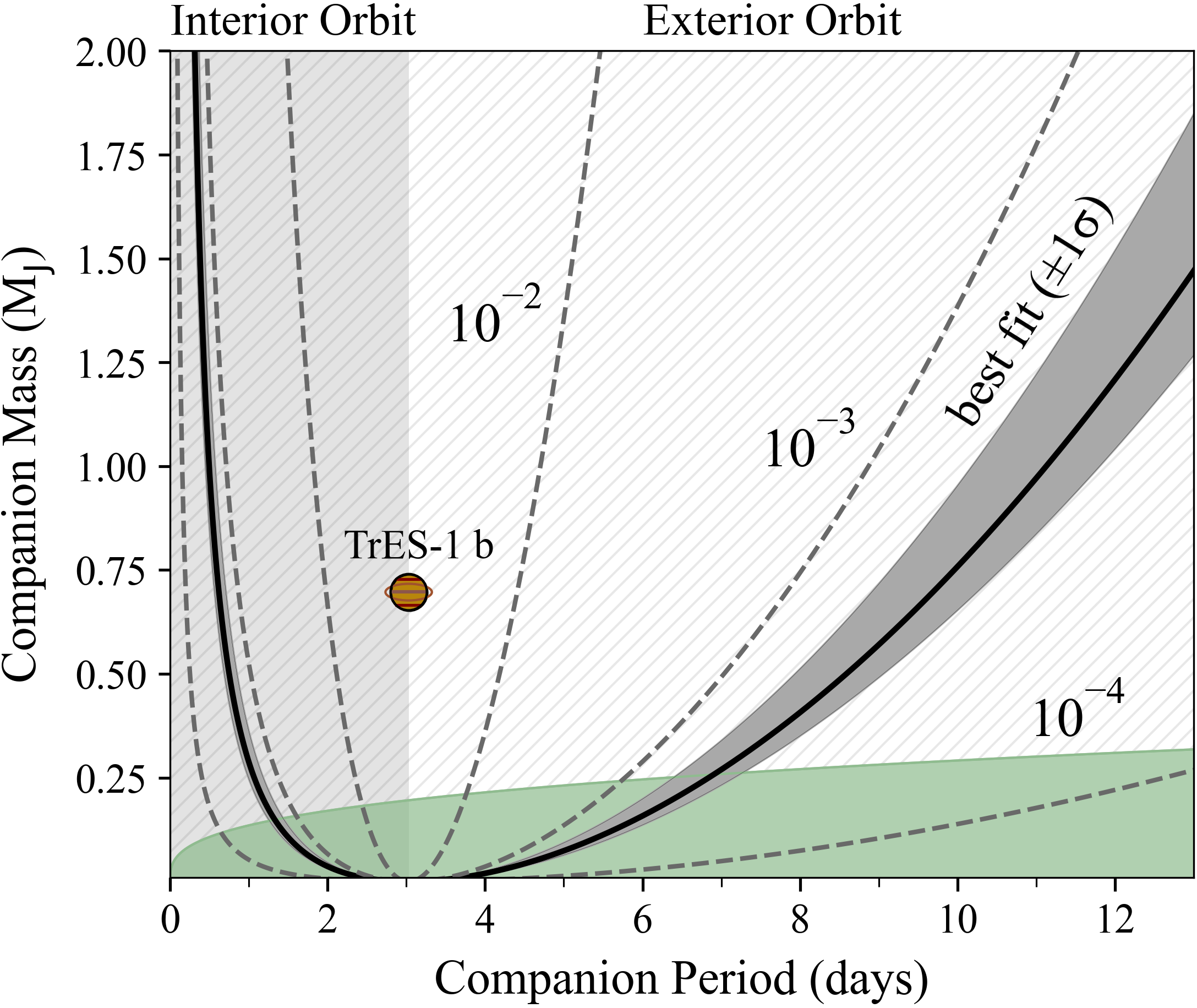}
        \caption{Mass vs. orbital period for a hypothetical companion planet that is driving apsidal precession. The solid black line and surrounding grey-shaded region correspond to the best-fit precession rate and 68\% credible interval, $\dot{\omega} = 0.00058\,^{+\,0.00016}_{-0.00009}$ ${\rm rad\,E^{-1}}$. The dashed grey lines show additional precession rates for comparison ($10^{-2}$, $10^{-3}$, and $10^{-4}$  ${\rm rad\,E^{-1}}$). The location of TrES-1\,b is marked, dividing the plot into interior and exterior orbits. The green shaded area shows the region where the RV amplitude, $K_c$, is within the residuals of the TrES-1\,b orbit fit (i.e. $K_c\ll30$, see Figure\>\ref{fig:rv_planet_b}), which rules out the remaining parameter space (hatched).}
        \label{fig:perturber_zoomed}
        \vspace{0.5cm}
    \end{minipage}
    \begin{minipage}{1.0\columnwidth}
        \hspace*{-0.6cm}\includegraphics[width=1.1\textwidth]{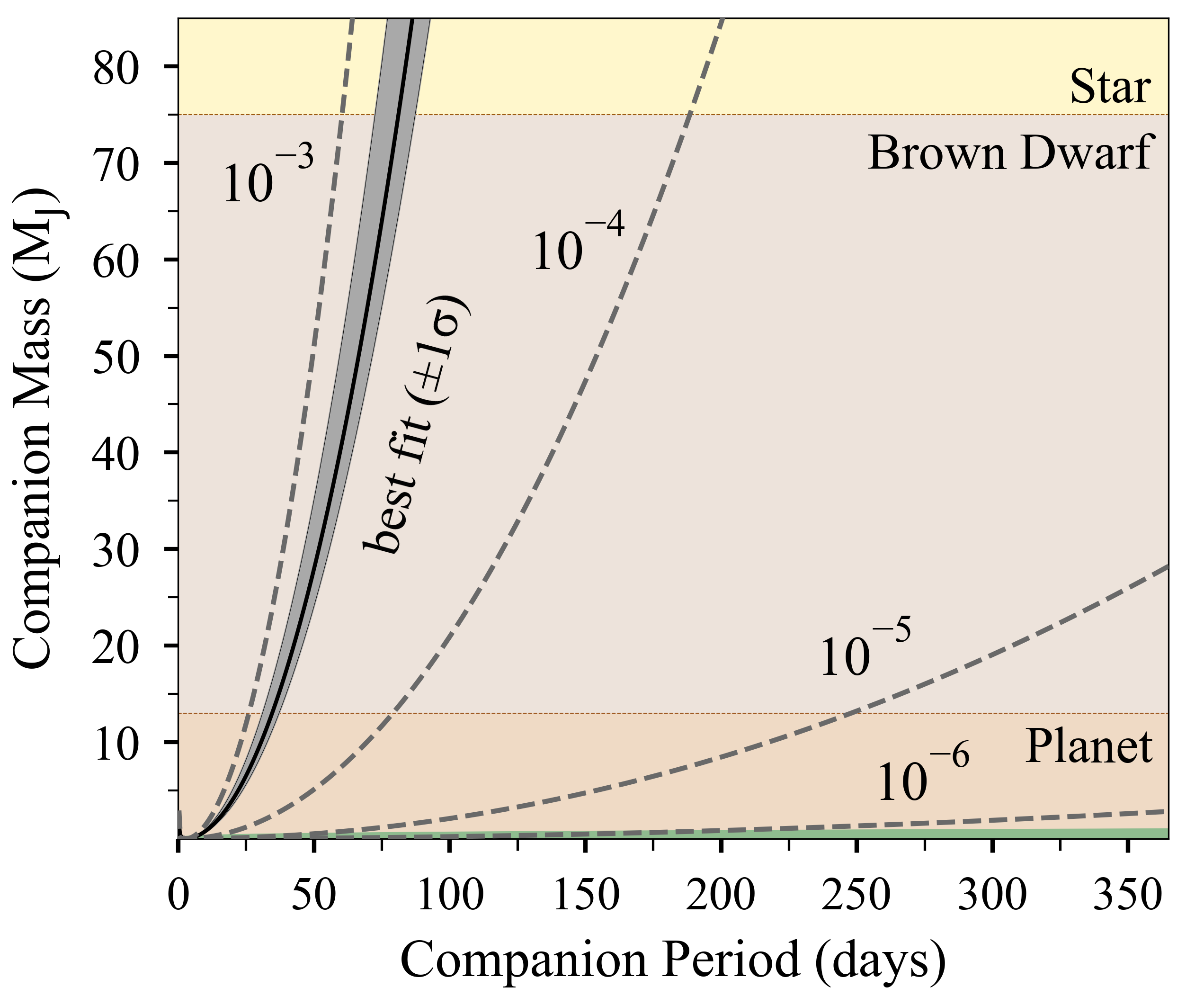}
        \caption{Similar to Figure\>\ref{fig:perturber_big}, but extended to include orbital periods up to one year. The three mass categories -- planet, brown dwarf, and star -- are illustrated along the y--axis. The green shaded region again denotes the RV detection limit, same as in Figure\>\ref{fig:perturber_big}.}
        \label{fig:perturber_big}
    \end{minipage}
\end{figure}

\subsubsection{Rotation}\label{sec:rotation}
Apsidal precession due to rotational flattening (oblateness) arises from the quadrupole term in the gravitational potential, parameterized by the Love number $k_2$.\footnote{The Love number represents how centrally condensed the body is. The lower the $k_2$, the more centrally condensed the planetary interior structure, which in turn leads to a slower precession rate. The theoretical upper limit of $k_2$ is $3/2$, which corresponds to a uniform density fluid object \citep{Lissauer2019}.} This rotational precession rate is expressed as \citep{Ragozzine2009}
\begin{equation}
    \begin{aligned}
     \dot{\omega}_{\rm rot} &= \dot{\omega}_{\rm rot,p} + \dot{\omega}_{\rm rot,\star}\,, \\
    &= \frac{k_{2_p}}{2} \left(\frac{\dot{\theta}_p}{a}\right)^2 \frac{n R_p^5}{G M_p}g_2(e), \\ 
    &+ \frac{k_{2_\star}}{2} \left(\frac{\dot{\theta}_\star}{a}\right)^2 \frac{n R_\star^5}{G M_\star}g_2(e),
    \end{aligned}
    \label{eq:rot}
\end{equation}
where,
\begin{equation*}
    g_2(e) = (1-e^2)^{-2}.
\end{equation*}
The rotation frequencies of the planet and star are $\dot{\theta}_p$ and $\dot{\theta}_\star$, respectively. For TrES-1, the rotational period is approximately $40.2 \pm 0.1$ days \citep{Dittmann2009}, while the rotation rate of TrES-1\,b is unknown. Assuming synchronous rotation (tidal locking), we adopt a rotation period of $3.03$ days for the planet. Using common but approximate Love numbers, we assume $k_{2p} = 0.3$ for the planet and $k_{2\star} = 0.03$ for the star \citep{Ragozzine2009}.

Together, these assumptions, observations, and model results yield
\begin{align*}
    \dot{\omega}_{\rm rot,p} &\approx 3 \times 10^{-3} \,^\circ\,{\rm yr}^{-1},~{\rm and} \\
    \dot{\omega}_{\rm rot,\star} &\approx 3 \times 10^{-5} \,^\circ\,{\rm yr}^{-1}.
\end{align*}

If we assume that the best-fit precession rate is dominated by either of these effects, inverting Equations \ref{eq:rot} yields nonphysical Love numbers for both the planet ($k_{2p} = 377$) and the star ($k_{2_\star} = 4034$), exceeding the theoretical upper limit of $3/2$. We find that no reasonable combination of stellar or planetary rotation periods and Love numbers can reproduce the observed apsidal precession rate.

\subsubsection{Tidal Bulges}\label{sec:tides}
Tidal bulges -- an ellipsoidal distortion -- are induced in both the planet and host star by their close proximity, another source of apsidal precession. The tidal precession rate depends on the internal density distribution of the planet, which is again parameterized by the $k_2$ Love number. The precession rate can be expressed as \citep{Ragozzine2009}
\begin{equation}
    \begin{aligned}
    \dot{\omega}_{\rm tide} &= \dot{\omega}_{\rm tide,p} + \dot{\omega}_{\rm tide,\star}, \\
    &= \frac{15}{2}\,k_{2p}\,n\left(\frac{R_p}{a}\right)^5\left(\frac{M_\star}{M_p}\right)f_2(e),~{\rm and}\\ 
    &+ \frac{15}{2}\,k_{2\star}\,n\left(\frac{R_\star}{a}\right)^5\left(\frac{M_p}{M_\star}\right)f_2(e),
    \end{aligned}
    \label{eq:tide}
\end{equation}
where,
\begin{equation*}
    f_2(e) = (1-e^2)^{-5} \left(1 + \frac{3}{2}e^2 + \frac{1}{8}e^4\right).
\end{equation*}
With the same assumptions for $k_{2p}$ and $k_{2\star}$, we find
\begin{align*}
    \dot{\omega}_{\rm tide,p} &\approx 5 \times 10^{-2} \,^\circ\,{\rm yr}^{-1},~{\rm and} \\
    \dot{\omega}_{\rm tide,\star} &\approx 6 \times 10^{-5} \,^\circ\,{\rm yr}^{-1}.
\end{align*}
Similar to rotation-induced precession, reproducing the observed rate would require unrealistically large Love numbers ($k_{2p} = 25$ and $k_{2_\star} = 2016$), and no reasonable value for either parameter can produce a tidal precession rate sufficient to explain the observations.

\subsection{Planet-Planet Precession} \label{sec:planet_precession}
Gravitational interactions between planets can drive apsidal precession, as seen in the Solar System. Following \citet{Heyl2007}, the precession rate induced by a companion, either interior or exterior to TrES-1\,b, is
\begin{eqnarray}
    \begin{aligned}
    \dot{\omega} &= \frac{m_2}{M_*} \frac{\alpha}{(\alpha+1)(\alpha-1)^2} \\ 
    & \times\left[\left(\alpha^2+1\right) \mathcal{E}\left(\frac{2 \alpha^{1 / 2}}{\alpha+1}\right)-(\alpha-1)^2 \,\,\mathcal{K}\left(\frac{2 \alpha^{1 / 2}}{\alpha+1}\right)\right],
    \end{aligned}
\end{eqnarray}
where $m_2$ is the mass of the perturbing planet, $\alpha = a_1/a_2$ is the ratio of TrES-1\,b's semi-major axis ($a_1$) to the companion's semi-major axis ($a_2$), and $\mathcal{K}$ and $\mathcal{E}$ are the complete elliptic integrals of the first and second kind, respectively. This equation assumes that (1) the companion planet is in a circular, coplanar, and nonresonant orbit, (2) $m_2 \ll M_\star$, and (3) the companion's gravitational potential can be approximated by spreading the companion's mass evenly along its orbit into a ring.

Figures \ref{fig:perturber_zoomed} and \ref{fig:perturber_big} illustrate the perturbing planet's mass-period parameter space with curves representing a companion planet that could drive the apsidal precession of TrES-1\,b's orbit. The companion mass increases dramatically with its orbital period, such that there is only a small region in which it could evade detection in the RV measurements (i.e. $K_2<30$ m/s, see Figure\>\ref{fig:rv_planet_b}). For a planetary companion to drive apsidal precession and remain undetected in the RV data, it would need to have a mass $\lesssim0.25$ M$_{\rm J}$ and an orbital period $\lesssim 7$ days for this system (Figure\>\ref{fig:perturber_zoomed}), while remaining dynamically stable.

No evidence of additional planets in the TrES-1 system have been found via short-term TTVs \citep{steffen_t1,Hrudkova2009,Raetz2009,Baluev2015,Rabus2008,rabus_t1_2009,Yeung2022}, even though such an object would likely have been detected. 
In particular, both \citet{steffen_t1} and \citet{rabus_t1_2009} exhaustively searched the 2:1 and 3:2 mean motion resonances (MMRs) and concluded that companions at these orbits with masses above one Earth mass are very unlikely. Moreover, the tentative TrES-1\,c revealed in the RV data (Section \ref{sec:results_rv}) is unable to drive the necessary precession.

\subsection{Orbital Decay}\label{sec:orbital_decay}
In the classical framework of gravitational tides developed by \citet{Darwin1908}, orbital decay occurs when a planet's orbital frequency exceeds the rotational frequency of its host star. This frequency mismatch causes the star’s tidal bulge to lag behind the planet, generating a net torque that transfers angular momentum from the planet’s orbit to the star’s rotation. These forces, arising from misaligned tidal bulges, are known as ``equilibrium tides.''

Orbital decay is driven by energy dissipation due to tides in the star and, if the planet remains in a non-synchronous rotation state, within the planet itself. The efficiency of this energy loss is typically quantified by the \textit{modified} tidal quality factor, $Q^{'}=3Q/2k_2$, which is inversely proportional to the energy dissipated per cycle \citep{Goldreich1966}.

We adopt tidal evolution models based on the ``viscous'' approach to equilibrium tides \citep[e.g.,][]{Hut1981,Levrard2007,Correia2010,Leconte2010} and evaluate orbital decay due to tides in both the star (Section \ref{sec:star_tides}) and the planet (Section \ref{sec:planet_tides}) separately. For clarity and completeness, Appendix \ref{appendix:tides_general_equations} summarizes the general equations for the period decay rate, $dP/dt$, which remain exact for any eccentricity, rotation frequency, and obliquity.

\begin{figure}
    \begin{minipage}[c]{1.0\columnwidth}
        \hspace*{-0.5cm}\includegraphics[width=1.05\textwidth]{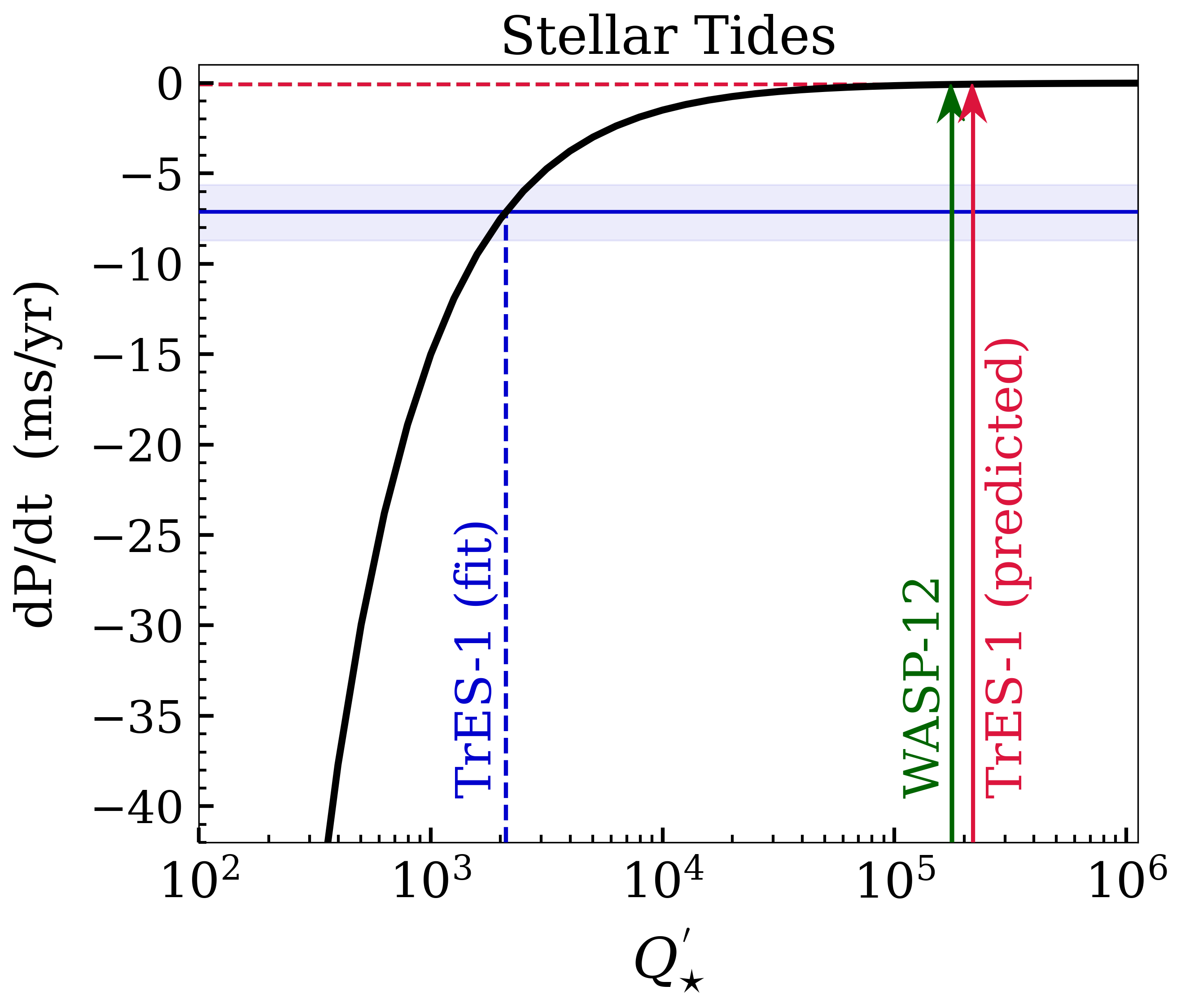}
        \caption{The expected orbital period derivative for TrES-1\,b under the assumption that stellar tides dominate its orbital evolution, as a function of the modified stellar quality factor $Q^{'}_\star$. The solid blue horizontal line represents the best-fit orbital decay rate for TrES-1\,b, with the shaded region indicating the corresponding 68\% credible interval. Three specific values of $Q^{'}_\star$ are highlighted: (1) the value derived from the best-fit orbital decay rate using Equation~\ref{eq:pdot_tides_star_2}, (2) the predicted value from the tidal forcing period of the system using Equation~\ref{eq:Q_predict}, and (3) the value inferred for WASP-12, a hot Jupiter decaying at $\sim30$ ${\rm ms\,yr^{-1}}$, from \citet{Wong2022}.}
        \label{fig:quality_factor}
    \end{minipage}

    \begin{minipage}[c]{1.0\columnwidth}
        \hspace*{-0.4cm}\includegraphics[width=1.04\textwidth]{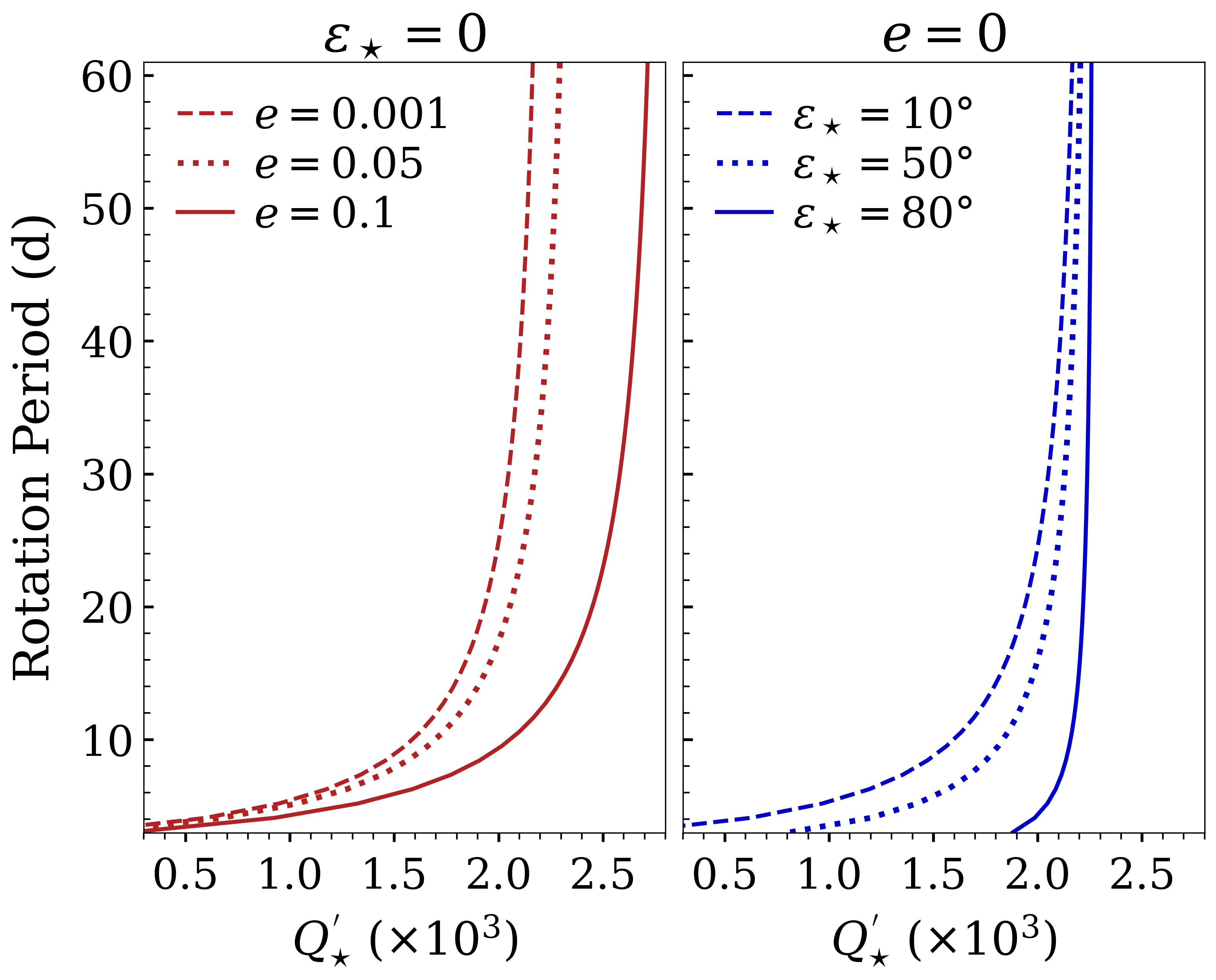}
        \caption{Stellar rotation period as a function of the modified stellar quality factor, $Q_\star^{'}$, for a scenario in which stellar tides drive orbital decay at a rate of $-7.1$ ${\rm ms\,yr^{-1}}$. The calculations use the exact expression for decay due to tides in the star (Equation \ref{eq:pdot_tides_star}) for different combinations of stellar obliquity ($\varepsilon_\star$) and orbital eccentricity ($e$). There is no parameter space in which $Q_\star^{'}$ falls within the empirically expected range of $10^5 < Q^{'}_\star < 10^8$ \citep{Penev2018}. \textbf{Left:} The required $Q_\star^{'}$ for different eccentricities when the stellar obliquity is zero. \textbf{Right:} The required $Q_\star^{'}$ for different obliquities when the orbital eccentricity is zero.}
        \label{fig:quality_factor2}
    \end{minipage}
\end{figure}

\subsubsection{Orbital Decay due to Tides in the Star} \label{sec:star_tides}
The rate of orbital decay due to the tidal response of the host star depends on its spin frequency, $\dot{\theta}_\star$, obliquity, $\varepsilon_\star$, and modified quality factor, $Q_\star^{'}$, as well as the orbital eccentricity. The full expression for the orbital period derivative is given in Equation~\ref{eq:pdot_tides_star} (Appendix~\ref{appendix:tides_general_equations}), but under the assumptions $e \approx 0$ and $\varepsilon_\star \approx 0$, it simplifies to
\begin{equation} \label{eq:pdot_tides_star_2}
    \frac{dP}{dt}_{\rm (star)} = -\frac{27\pi}{Q_\star^{'}} \left(\frac{M_p}{M_\star}\right) \left(\frac{R_\star}{a}\right)^5 \left[1 - \frac{\dot{\theta}_\star}{n}\right].
\end{equation}

For TrES-1, inverting this equation yields $Q^{'}_\star \approx 2.1 \times 10^{3}$ for the best-fit orbital decay rate of $-7.1$ ${\rm ms\,yr^{-1}}$ and a stellar rotation period of $40.2$ days \citep{Dittmann2009},\footnote{If the host star's rotation period is unknown, Equation~\ref{eq:pdot_tides_star_2} can be further simplified by assuming $\dot{\theta}_\star \ll n$, in which case the bracketed term approaches unity.} which is two orders of magnitude smaller than the derived value for WASP-12 ($Q^{'}_\star = 1.75 \times 10^5$, \citealt{Yee2019}).

This value is well below the commonly cited range of $10^5 < Q^{'}_\star < 10^8$, which is supported by multiple lines of empirical evidence -- including constraints from tidal circularization timescales in binary star systems \citep{Meibom2005, Milliman2014} and close-in giant planet orbits \citep{Jackson2008, Husnoo2012, Bonomo2017}, observational limits on tidal decay rates inferred from transiting planet properties \citep{Penev2012}, and observed rotation rates of HJ host stars evaluated in the context of tidal spin-up \citep{Penev2018}.

As an alternative test, we apply the empirical model for $Q_\star^{'}$ from \citet{Penev2018},\footnote{This model is based on a sample of exoplanet systems with $M_p > 0.1$ M$_{\text{J}}$, $P < 3.5$ days, and $T_{\rm eff,\star} < 6100$\,K, a parameter space that includes TrES-1\,b.} which relates $Q_\star^{'}$ to the tidal forcing period, $P_{\rm tide}$,
\begin{equation} \label{eq:Q_predict}
    Q^{'}_\star = \max\left[{\frac{10^6}{(P_{\rm tide}/\,{\rm day})^{3.1}}},\,{10^5}\right],
\end{equation}
where
\begin{equation*} \label{eq:p_tide}
    P_{\rm tide} = \frac{1}{{2\left(P_{\rm orb}^{-1} - P_{\rm rot}^{-1}\right)}}.
\end{equation*}

Using $P_{\text{orb}}=3.03$ days and the stellar rotation period of $P_{\text{rot}} = 40.2$ days \citep{Dittmann2009}, Equation\>\ref{eq:Q_predict} yields $Q_\star^{'} \approx 2 \times 10^5$. While this value aligns with theoretical expectations, it corresponds to an orbital decay rate of only $-0.07$ ${\rm ms\,yr^{-1}}$, much smaller than the observed rate. As demonstrated by Figure\>\ref{fig:quality_factor}, even the lower bound of theoretical expectations for $Q^{'}_\star$ fails to produce orbital decay that could be detected on decade timescales.

We note that the above analysis assumes $e = 0$ and $\varepsilon_\star = 0$, but the eclipse timing data indicates a nonzero eccentricity and the \textit{projected} stellar obliquity has been measured as $\lambda = 30\pm21$ degrees \citep{Narita2007}. Though $\lambda$ is not the true obliquity, it implies a nonzero $\varepsilon_\star$ value. However, as illustrated in Figure\>\ref{fig:quality_factor2}, no combination of $e$ and $\varepsilon_\star$ enhances tidal dissipation in the star enough to yield a physically reasonable $Q^{'}_\star$. Altogether, this suggests that stellar tides are unlikely to drive the detected secular evolution of TrES-1\,b’s orbit.

\begin{figure*}
    \begin{minipage}[c]{1.0\columnwidth}
        \hspace*{-0.3cm}\includegraphics[width=1.05\textwidth]{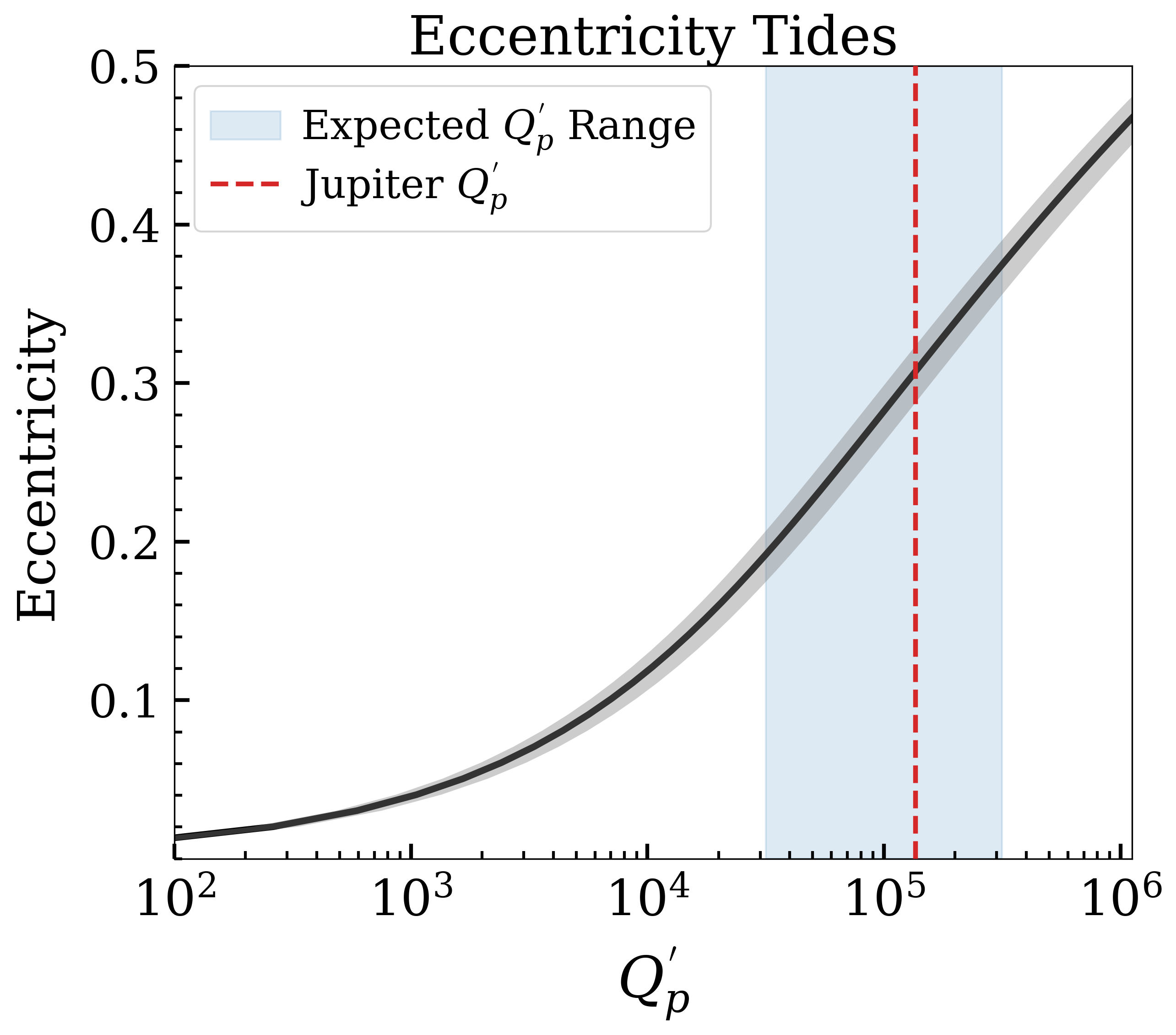}
        \caption{Required orbit eccentricity ($e$) as a function of the modified tidal quality factor ($Q_p^{'}$) for TrES-1\,b, assuming eccentricity tides are the dominant mechanism driving orbital decay. The black curve and surrounding gray region show the eccentricity values needed to match the best-fit decay rate of $-7.1^{+\,1.5}_{-1.6}$ ms yr$^{-1}$. The shaded blue region represents the theoretically expected range of $Q_p^{'}$ for HJs \citep{Mahmud2023}, while the red dashed line marks the estimated value for Jupiter \citep{Lainey2009}. The fact that the required eccentricity ($e \gtrsim 0.2$) falls well outside observational constraints suggests that eccentricity tides cannot drive orbital decay.}
        \label{fig:Q_eccentricity}
    \end{minipage}
    \hfill
    \vspace{0.4cm}
    \begin{minipage}[c]{1.0\columnwidth}
        \vspace{-0.1cm}
        \hspace*{-0.3cm}\includegraphics[width=1.05\textwidth]{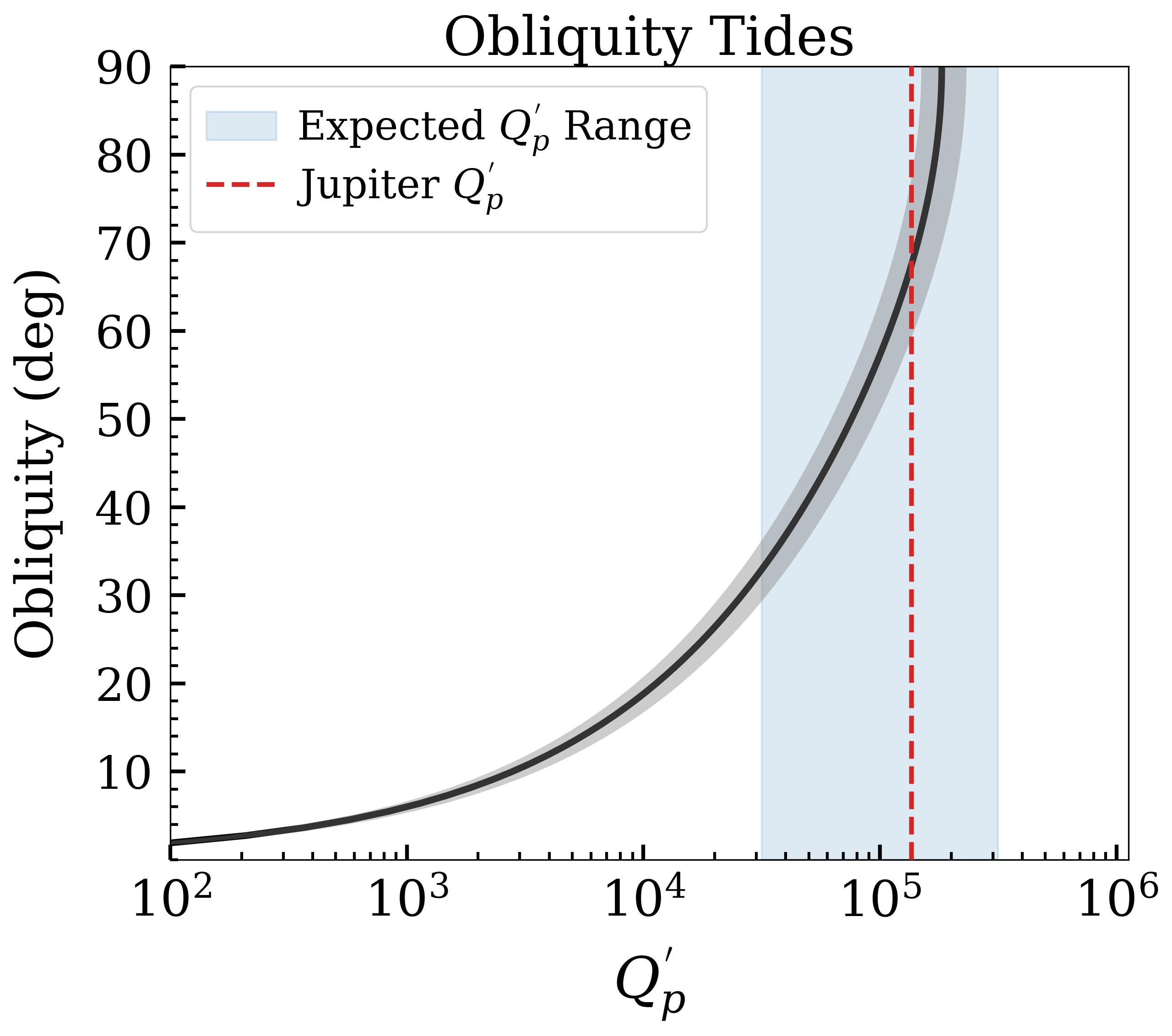}
        \caption{Required planetary obliquity ($\varepsilon_p$) as a function of the modified tidal quality factor ($Q_p^{'}$) for TrES-1\,b, assuming obliquity tides are the dominant mechanism driving orbital decay. The black curve and surrounding gray region show the obliquity values needed to match the best-fit decay rate of $-7.1^{+\,1.5}_{-1.6}$ ms yr$^{-1}$. The shaded blue region represents the theoretically expected range of $Q_p^{'}$ for HJs \citep{Mahmud2023}, while the red dashed line marks the estimated value for Jupiter \citep{Lainey2009}. Obliquity values $\varepsilon_p \gtrsim 30^\circ$ yield $Q_p^{'}$ estimates that are consistent with theoretical expectations.}
        \label{fig:Q_obliquity}
    \end{minipage}
\end{figure*}

\subsubsection{Orbital Decay due to Tides in the Planet} \label{sec:planet_tides}
The tidal response of the planet may contribute to orbital decay if the planet has a nonzero eccentricity and/or obliquity ($\varepsilon_p$), meaning it is not in a perfectly synchronous rotation state. Our analysis of TrES-1\,b suggests that a small but nonzero eccentricity is required to explain the observed offset in the eclipse midtimes (Figure \ref{fig:o-c}), motivating an investigation into the role of energy dissipation within TrES-1\,b.

The full expression for the orbital period decay rate due to planetary tides is given by Equation \ref{eq:pdot_tides_planet} (Appendix \ref{appendix:tides_general_equations}). This equation assumes that the planet’s spin frequency has settled into an equilibrium state determined by its eccentricity and obliquity. In the following sections, we separately evaluate the contributions from “eccentricity tides” (where $\varepsilon_p = 0$) and “obliquity tides” (where $e = 0$), while noting that both mechanisms could, in principle, act simultaneously.

\paragraph{\textbf{Eccentricity tides}} When $\varepsilon_p = 0$, the period decay rate due to planetary tides is given by:
\begin{equation} \label{eq:pdot_tides_eccentricity}
    \frac{dP}{dt}_{\rm (ecc)} = -\frac{27\pi}{Q_p^{'}} \left(\frac{M_\star}{M_p}\right) \left(\frac{R_p}{a}\right)^5 \left[f(e) - \frac{g^2(e)}{h(e)}\right],
\end{equation}
where $f(e)$, $g(e)$, and $h(e)$ are functions of eccentricity defined in Equations \ref{eq:tides_f}, \ref{eq:tides_g}, and \ref{eq:tides_h} (Appendix \ref{appendix:tides_general_equations}).

Assuming that all tidal dissipation occurs within the planet, we invert Equation \ref{eq:pdot_tides_eccentricity} and use the best-fit decay rate along with the measured eccentricity ($e = 0.00126$, Appendix \ref{appendix:results_table}) to estimate a modified planetary quality factor of $Q_p^{'} \approx 1$. This value is physically implausible. For HJs, estimates place their modified tidal quality factors in the range $10^{4.5} < Q_p^{'} < 10^{5.5}$ \citep{Mahmud2023}, although some studies suggest even higher values (indicating less efficient energy dissipation).\footnote{See Section 5 of \citet{Mahmud2023} for a detailed summary of empirical constraints on $Q_p^{'}$.}

As shown in Figure \ref{fig:Q_eccentricity}, an eccentricity of $e \gtrsim 0.2$ would be required for $Q_p^{'}$ to fall within the expected range. However, such a high eccentricity is strongly ruled out by observations, leading us to conclude that eccentricity tides cannot explain the detected variations in TrES-1\,b's orbit.

\paragraph{\textbf{Obliquity tides}}
When $e = 0$, the period decay rate due to tides in the planet is:
\begin{equation} \label{eq:pdot_tides_obliquity}
    \frac{dP}{dt}_{\rm (obl)} = -\frac{27\pi}{Q_p^{'}} \left(\frac{M_\star}{M_p}\right) \left(\frac{R_p}{a}\right)^5 \left[1 - \frac{2 \cos^2{\varepsilon_p}}{(1 + \cos^2{\varepsilon_p})}\right].
\end{equation}

We find that obliquity tides \textit{can} account for the observed evolution of TrES-1\,b's orbit if the planetary obliquity is $\varepsilon_p \gtrsim 30^\circ$ (Figure \ref{fig:Q_obliquity}). While there is no direct measurement of TrES-1\,b’s true obliquity, \citet{Narita2007} suggest that the planet rotates in a prograde direction, implying a value in the range of $0^\circ < \varepsilon_p < 90^\circ$.

For obliquities in the range $30^\circ < \varepsilon_p < 90^\circ$, the best-fit orbital decay rate of $-7.1^{+\,1.5}_{-1.6}$ ms yr$^{-1}$ corresponds to modified quality factors in the range $10^{4.4} < Q_p^{'} < 10^{5.3}$ (Figure \ref{fig:Q_obliquity}) -- values that are consistent with theoretical expectations for hot Jupiters. Obliquity tides have similarly been proposed as an explanation for the rapid orbital decay of WASP-12\,b, where \citet{Millholland2018} found that an obliquity exceeding $50^\circ$ could account for the observed decay rate.

The strong agreement between the best-fit decay model for TrES-1\,b and theoretical predictions for $Q_p^{'}$ is encouraging.  However, maintaining a nonzero planetary obliquity over long timescales requires an additional mechanism to counteract tidal damping. One possibility is that the planet is trapped in a spin-orbit resonance due to perturbations from an undetected outer companion \citep{Millholland2020}. While investigating this scenario is beyond the scope of this study, we encourage future work in this direction.
 
\subsection{Line-of-Sight Acceleration} \label{sec:line_of_sight}
Rather than true orbital evolution, quadratic timing changes could potentially be explained by a distant perturber with an orbital period much longer than the observational baseline ($18$ years of transits). In this model, there is a variation in light travel time caused by a steady acceleration along the line-of-sight, 
which may result in an \textit{apparent} change in the orbital period of TrES-1\,b, the rate of which is related to the radial acceleration, $\dot{\gamma}$, by
\begin{equation}\label{eq:romer}
    \dot{P} = \frac{\dot{\gamma} P}{c},
\end{equation}  
where $P$ is the orbital period of TrES-1\,b, and $c$ is the speed of light. For orbital decay, the line-of-sight acceleration must be negative, indicating that the star is moving toward us. 

Using Equation\>\ref{eq:romer}, the best-fit decay rate of $-7.1~\rm ms~yr^{-1}$ corresponds to $\dot{\gamma} = -0.022~\rm m~s^{-1}~day^{-1}$. 
Such an acceleration would cause a noticeable linear trend in the data, which is not seen. 
However, given the shorter RV baseline (12 years), wide gaps, and large uncertainties in early observations (Figure~\ref{fig:rv_planet_b}), we explore this situation further below.   

\subsubsection{Acceleration by a Stellar-Mass Object} \label{sec:los_binary_star}
High-resolution optical imaging surveys \citep{Faedi2013,Adams2013,Ngo2016} have identified three potential stellar companions. The closest is at $2.340{''} \pm 0.01{''}$, but it is too faint for a reliable flux measurement with the given observations. The second, at $4.940{''} \pm 0.002{''}$, is likely a main-sequence star with an estimated mass of $\sim0.15$ M$_{\odot}$. 
The third, at $6.355{''} \pm 0.002{''}$, is most likely a background object.

Using the framework of \citet{Torres1999}, the minimum mass of a stellar companion causing a radial acceleration is given by
\begin{equation}\label{eq:romer_star}
    M_B = 5.341 \times 10^{-6} \left(\frac{D}{\mathrm{pc}}\>\frac{\rho}{\mathrm{arcsec}}\right)^{2} \left| \frac{\dot{\gamma}}{\mathrm{m\,s^{-1}\,day^{-1}}} \right| \Phi,
\end{equation}
where $\rho$ is the angular separation in arcseconds, $D$ is the system distance, $\dot{\gamma}$ is the radial acceleration, and $\Phi$ is a unitless parameter that is dependent on the companion orbit's eccentricity, argument of pericenter, and inclination. Since the latter are unconstrained, we adopt the minimum value of $\Phi = \frac{3\sqrt{3}}{2}$ \citep{Torres1999}. For a radial acceleration of $\dot{\gamma} = -0.022~\rm m~s^{-1}~day^{-1}$, the imaged companions would need to be unrealistically massive, with lower masses of $16$ M$_{\odot}$ and $69$ M$_{\odot}$ for the $2.340{''}$ and $4.940{''}$ objects, respectively. 

An alternative scenario combines the apsidal precession model, accounting for $-6.4~\rm ms~yr^{-1}$ (see Section\>\ref{sec:results}), with a residual light travel time effect of $-0.7~\rm ms~yr^{-1}$, to give the required acceleration $\dot{\gamma} = -0.0022~\rm m~s^{-1}~day^{-1}$. This smaller trend could easily remain undetected in the RV data. However, even in this case, the minimum companion masses are $1.5$ M$_\odot$ and $6.8$ M$_\odot$ for separations of $2.340{''}$ and $4.940{''}$, respectively. Though more plausible, these masses remain inconsistent with observed magnitude constraints, making it unlikely that these stellar companions cause any observed variations in TrES-1\,b's orbit.

\subsubsection{Acceleration by a Planetary-Mass Object} \label{sec:los_planet}
Following the framework outlined by \citet{Feng2015,Bouma2020}, we estimate a lower mass for the case of a planetary perturber using
\begin{equation}\label{eq:romer_planet}
    M_{c} \approx 5.99 \,{\rm M}_{\rm Jup} \left( \frac{\tau}{\mathrm{yr}} \right)^{4/3} 
    \left| \frac{\dot{\gamma}}{\mathrm{m\,s^{-1}\,day^{-1}}} \right| 
    \left( \frac{M_{\star}}{M_{\odot}} \right)^{2/3}.
\end{equation}

To determine an absolute minimum mass, Equation\>\ref{eq:romer_planet} is derived by assuming that the companion's orbit has an eccentricity of $e=0.5$, an argument of pericenter $\omega = \pi/2$, and an orbital period equal to $1.25\tau$, where $\tau$ is the observational baseline in years.\footnote{To understand why this gives the minimum mass, imagine a hypothetical RV signal for an eccentric planet, which resembles a sawtooth curve, such that the observations correspond only to the linear portion of signal (see Figure 1 of \citealt{Feng2015}). }

To match a quadratic timing model, we use the transit baseline of 18.4 years rather than the shorter 12-year radial velocity (RV) baseline. Again using $\dot{\gamma} = -0.022~\rm m~s^{-1}~day^{-1}$, the above equation yields $M_c > 5.9$ M$_{\mathrm{J}}$, corresponding to an orbital period of approximately $P_c = 1.25\times \tau = 23$ years ($\sim7.7$ au). As before, if we instead consider a combination of the apsidal precession model with a residual light travel time effect, corresponding to $\dot{\gamma} = -0.0022~\rm m~s^{-1}~day^{-1}$, the limit becomes $M_c > 0.59$ M$_{\mathrm{J}}$. While such companions are not ruled out, we emphasize the current RV data do not present evidence for their existence, either. Nonetheless, this possibility should be kept in mind, as extending the RV baseline is essential to confirm or rule out such perturbers. 

\subsection{Outer Planetary Companion on an Eccentric Orbit} \label{sec:outer_planet}
In Section\>\ref{sec:results_rv} we found that the residuals of the RV data (i.e., with the TrES-1\,b fit removed, Figure\>\ref{fig:rv_planet_c}) show evidence for a potential planetary companion on a wide, eccentric orbit. While these orbital periods are shorter than the observational baseline and therefore cannot explain secular trends, we examine whether perturbations from these possible companions are significant enough to influence the orbit of TrES-1\,b. 

As a reminder, the periodogram of the residuals shows peaks at 3567, 1788, 1383, and 1194 days (Figure\>\ref{fig:periodogram}). Multiple models were fit with priors centered on each peak, and in all cases, the best-fit models favor highly eccentric orbits. The overall best-fit solution is (Table\>\ref{table:rv_results})
\begin{equation*}
    P_c=1200^{\,+\,26}_{\,-20}\,{\rm days},\>e_c=0.65^{\,+\,0.1}_{\,-0.1}\,,\>M_c\sin i_c \approx 0.36\,{\rm M}_{\mathrm{J}}.
\end{equation*}
The other peaks have their own best-fit orbital solutions, but the Bayesian evidence decreases with increasing period. For completeness, those solutions are:
\begin{equation*}
    \begin{aligned}
        P_c&=1390^{\,+\,19}_{\,-20}\,{\rm days},\>e_c=0.71^{\,+\,0.11}_{\,-0.13}\,,\>M_c\sin i_c \approx 0.42\,{\rm M}_{\mathrm{J}}, \\
        P_c&=1782^{\,+\,30}_{\,-27}\,{\rm days},\>e_c=0.82^{\,+\,0.08}_{\,-0.11}\,,\>M_c\sin i_c \approx 0.48\,{\rm M}_{\mathrm{J}}, \\
        P_c&=3597^{\,+\,41}_{\,-41}\,{\rm days},\>e_c=0.84^{\,+\,0.08}_{\,-0.10}\,,\>M_c\sin i_c \approx 0.79\,{\rm M}_{\mathrm{J}}.
    \end{aligned}
\end{equation*}

The planet-planet precession rates for the above companion solutions are negligible (see Figure\>\ref{fig:perturber_big}), so we focus on two effects: (1) changes in light travel time due to the reflex motion of the inner star-planet binary and (2) gravitational interactions.

\subsubsection{Light Travel-Time Variations}
A massive companion on a wide orbit can cause transit timing variations by inducing light travel time changes due to the reflex motion of the inner binary (star and planet). This light travel time (LTT) mechanism is the same as that described in Section\>\ref{sec:line_of_sight}, except that here, the orbital period is shorter than the observational baseline, so the entire periodic effect may be captured in the data. 

The full extent of the light travel time variation occurs when the star goes from its closest to farthest point relative to the observer. As such, we can express the semi amplitude of the LTT variations, $\Delta t$, as 
\begin{equation}
	\Delta t = \frac{M_c}{M_\star} \frac{a_c}{c},
\end{equation}
where $a_c$ is the semi-major axis of the companion’s orbit, $M_c$ its mass, $M_\star$ the stellar mass, and $c$ the speed of light. 

For the companion solutions with orbital periods of $1200$, $1390$, $1782$, and $3597$ days, the corresponding semi amplitudes are $0.42$ s, $0.53$ s, $0.71$ s, and $1.9$ seconds. These values are too small to be detectable and cannot contribute to the observed variations in TrES-1\,b’s orbit. 

\subsubsection{Gravitational Interactions} \label{sec:companion_tide_ttv}
Gravitational interactions, or tidal forces, between the planetary bodies can induce TTVs independent of light travel time effects. Following the framework of \citet{Agol2005}, the inner planet samples the gravitational potential of the outer planet, leading to TTVs with the same period as the perturber's orbit. The semi amplitude of the variations is given by $\Delta t = \sqrt{2}\,\sigma_{\rm TTV}$, where $\sigma_{\rm TTV}$ is the standard deviation, expressed as 
\begin{equation}\label{eq:companion_ttv_gravitational}
	\sigma_{\rm TTV} = \frac{3 A \, e_c}{\sqrt{2}\left(1-e_c^2\right)^{3/2}}
	\left[1 - \frac{3}{16} e_c^2 - \frac{47}{1296} e_c^4 - \frac{413}{27648} e_c^6 \right]^{1/2}
\end{equation}
where,
\begin{equation*}
	A = \frac{M_c}{2\pi\left(M_\star+M_p\right)} \frac{P^2}{P_c}.
\end{equation*}
Here, $P_c$, $e_c$, and $M_c$ represent the companion’s orbital period, eccentricity, and mass, respectively, while $P$ and $M_p$ denote the period and mass of TrES-1\,b.

Using Equation\>\ref{eq:companion_ttv_gravitational}, we calculate the TTV semi amplitudes for the proposed companion solutions. These are $0.18$ s, $0.24$ s, $0.44$ s, and $0.43$ s for periods of $1200$, $1390$, $1782$, and $3597$ days, respectively. These variations are also too small to be observable in current data sets and therefore cannot explain the observed variations in TrES-1\,b’s orbit. 

\subsection{Systemic Motion}\label{sec:proper_motion}
The movement of a star-planet system through space, relative to the solar system, could give rise to apparent changes in the given planet's orbit. This systemic motion is partially constrained by the proper motion on the sky, which for TrES-1 is available in the Gaia DR3 catalogue (Table\>\ref{tab:properties}). Using these data, we consider two effects.

First, as a star-planet system moves through space, a nonzero transverse velocity will cause the radial velocity component to change over time. In turn, this causes a steady change in the observed orbital period, denoted $\dot{P}_{\rm Shk}$. This phenomenon, known as the Shklovskii Effect, was first identified in pulsar timing studies \citep{Shklovskii1970}. 
For exoplanet studies, \citet{Rafikov2009} derives an convenient expression for $\dot{P}_{\rm Shk}$, given by
\begin{equation} \label{eq:shklovskii}
    \dot{P}_{\textup{Shk}} = 20\left(\frac{\mu}{100\,{\rm mas\,yr}^{-1}}\right)^2\frac{D}{100\,{\rm pc}}\frac{P}{3\,\rm day}\,\mu {\rm s\,yr}^{-1},
\end{equation} 
where $\mu = |\vec{\mu}|$ is the magnitude of the proper motion vector, $D$ is the distance to the system, and $P$ is the orbital period. Using the Gaia DR3 parameters for TrES-1 (Table\>\ref{tab:properties}), we calculate $\dot{P}_{\rm Shk} = 0.005$ ${\rm ms\,yr^{-1}}$, which is far below detection limits over the 18-year observation baseline of TrES-1\,b.

The second effect we consider is the \textit{apparent} apsidal precession of the orbit, $\dot{\omega}_\mu$, due to its angular reorientation in the sky. \citet{Rafikov2009} further derives the rate of this precession as
\begin{equation} \label{eq:dwdt_mu}
\dot{\omega}_\mu = -\mu\frac{\sin{\beta}}{\sin{i}},
\end{equation}
where $\beta$ represents the angle in the sky plane between the proper motion vector and the projection of the orbital angular momentum vector, and $i$ is the line-of-sight orbital inclination. The precession rate is maximized when $\beta$ equals $90^\circ$ or $270^\circ$. 

For TrES-1\,b, with $i = 88.4^\circ$ \citep{Narita2007}, the maximum possible value is $\dot{\omega}_\mu = 1.5\times10^{-9}\,{\rm rad}\,{\rm E}^{-1} = 1.1\times10^{-5} \,^\circ\,{\rm yr}^{-1}$, which is too small to be observable. Thus, we conclude that systemic motion effects cannot account for the observed variations in the orbit of TrES-1\,b.

\subsection{Magnetic Activity}\label{sec:magnetic_activity}
Finally, we consider the possibility of stellar magnetic activity as the origin of the observed variations of TrES-1\,b's orbit. We consider two cases: (1) an apparent period variation due to shifts in transit timing from starspot occultations, and (2) a true period variation driven by the Applegate effect, in which magnetic activity influences the star's quadrupole moment.

\subsubsection{Starspot Occultations} \label{sec:starspots}

The apparent mid-time of a transit can be altered whenever the planet passes over or near a starspot \citep{Oshagh2013, MillerRicci2008}. 
Under rare circumstances, starspot occultations may even emulate TTVs that are typically associated with resonant planetary companions \citep{Ioannidis2016}.

Prior studies have identified starspots on the surface of TrES-1 \citep{Dittmann2009, rabus_t1_2009_hst}, and \citet{Sozzetti2006} indicate it is an active star. Additionally, its relatively slow, 40 day rotation period likely maintains a fairly stable magnetic field \citep{Olah2016}. 

However, regardless of stable magnetic activity, starspots themselves are not static features; they come and go over periods ranging from a few weeks to several months, and their positions and sizes vary throughout their lifetime. Additionally, \citet{Narita2007} constrained the sky-projected angle between the stellar spin axis and the orbital plane to be within $30 \pm 21$ degrees, rendering continuous occultations of the same starspot unlikely. Thus, the observed long-term variations in TrES-1\,b's orbital period over a span of $18$ years suggest that the starspot hypothesis is improbable, as the persistent and systematic nature of the observed variations point to alternative mechanisms at play. 

\subsubsection{The Applegate Effect} \label{sec:applegate}
The Applegate effect \citep{Applegate1987,Applegate1992} is a mechanism behind some period variations observed in eclipsing binary star systems. In this mechanism, magnetic activity within the star drives a transfer of angular momentum between layers of the core and envelope. This causes a variation in the oblateness of the star, which in turn changes the quadrupole moment. The resulting evolution of the gravitational potential drives a quasi-periodic variation in the planet's orbital period.

\citet{Watson2010} show that this effect may be significant for some HJs, even finding that the timing deviations for WASP-18 b are on the same order of those expected from orbital decay. However, TrES-1\,b is included in their sample, and they predict a timing difference of just $\Delta {\rm t} = $ $0.1$, $0.2$, and $0.8$ seconds for stellar magnetic activity cycle periods of $11$, $22$, and $50$ years, respectively. Given the measurement uncertainties in the TrES-1\,b data set, the \citet{Watson2010} result indicates that the Applegate effect would not be detectable for the TrES-1 system within the current baseline of observations.

\section{Conclusion}\label{sec:conclusion}
Our analysis of TrES-1\,b's transit, eclipse, and radial velocity data identifies apsidal precession as the most plausible explanation for the observed variations in its orbit, regardless of the dataset subset used (Section~\ref{sec:results_subsets}). Even when eclipse midtimes are excluded, the precession model is either indistinguishable from orbital decay (when considering transit timing alone) or provides a better fit when including radial velocity (RV) data. Furthermore, the orbital decay model remains a stronger fit than a constant-period orbit, strengthening the interpretation that there is some form of secular variation in the system, regardless of whether it represents genuine orbital evolution or an apparent effect. 

However, while apsidal precession best explains the data, the rapid precession rate cannot be easily explained without invoking an additional, undetected planet in close proximity to  TrES-1\,b (Section\>\ref{sec:planet_precession}). We constrain such a companion’s parameter space to a mass $\lesssim0.25$ M$_{\rm J}$ and an orbital period $\lesssim 7$ days (Figure~\ref{fig:perturber_zoomed}), but emphasize that, apart from the observed secular variations, there is no direct evidence of such a planet in the TrES-1 system. Nonetheless, we ruled out general relativity, stellar rotation, and tidal interactions as primary mechanisms for the observed precession (Section\>\ref{sec:apsidal_precession}), as well as apparent trends caused by systemic motion (Section\>\ref{sec:proper_motion}).

The orbital decay model, under equilibrium tidal theory, requires an unreasonably high tidal dissipation efficiency if the host star is driving the decay (Section~\ref{sec:orbital_decay}). Dynamical tides could enhance dissipation within the star, but such mechanisms have struggled to explain the rapid decay rates of other systems like WASP-12\,b \citep{Bailey2019} and Kepler-1658\,b \citep{Barker2024}.

Instead, we find that tidal energy dissipation within the \textit{planet} can account for the best-fit orbital decay rate if its obliquity is nonzero. For planetary obliquities between $30^\circ < \varepsilon_p < 90^\circ$, the required tidal quality factors fall within the expected range for HJs (Figure~\ref{fig:Q_obliquity}), consistent with previous findings for WASP-12\,b \citep{Millholland2018}. This suggests that TrES-1\,b’s orbit could be decaying due to energy dissipation within the planet rather than the star, provided it is trapped in an asynchronous rotation state.

However, the obliquity-driven decay scenario challenges the conventional assumption that HJs are tidally locked into synchronous rotation. In the absence of external perturbations, synchronization and circularization timescales are expected to be on the order of millions of years \citep{Rasio1996}, much shorter than the system's age, making an asynchronously rotating TrES-1\,b surprising. If the planet is not synchronized, an additional torque -- such as perturbations from an outer companion -- may be sustaining its obliquity and eccentricity against tidal damping. Alternatively, recent work by \citet{Wazny2025} suggests that HJs may be driven away from synchronous rotation as a result of interactions between thermal atmospheric winds and the planet's magnetic field, a scenario which does not necessitate an additional body in the system.

Radial velocity evidence of a second planet in the TrES-1 system, TrES-1\,c (Section \ref{sec:results_rv}), further complicates the picture. While the wide, eccentric orbit of such a companion makes it unlikely to drive the observed variations of TrES-1\,b's orbit (Section \ref{sec:outer_planet}), its presence raises the possibility of additional dynamical interactions within the system.

We have ruled out other effects, including a line-of-sight acceleration from a distant companion planet or binary star (Section~\ref{sec:line_of_sight}), apparent variations due to systemic motion (Section~\ref{sec:proper_motion}), and stellar magnetic activity (Section~\ref{sec:magnetic_activity}).

Altogether, the observational evidence strongly suggests that there is a physical timing trend occurring in TrES-1\,b, with the apsidal precession model best describing the data. As the field of exoplanet timing progresses, it is clear that transit data and simple analytical models alone are insufficient. Understanding the intricate dynamics of systems like TrES-1\,b will require multifaceted approaches and continued investigation that integrates long-term transit monitoring with precise RV and eclipse observations, alongside N-body simulations.

\section*{Acknowledgments}
This study was made possible by observations from citizen scientists. In particular, a large portion of the transit light curve data was contributed by amateur astronomers and accessed through ExoClock \citep{kokori_exoclockI}, the Exoplanet Transit Database (ETD; \citealt{poddany_etd_2010}) through the VarAstro portal \citep{VarAstro}, and Exoplanet Watch \citep{Zellem2020, Pearson2022} through the American Association of Variable Star Observers (AAVSO; \citealt{aavso}). We thank the organizers of these programs and the citizen science community.

We gratefully acknowledge Šárka Dyčková and Vicenç Ferrando for contributing two transit observations to the ETD that were used in this study, taken at the Masaryk University Observatory (Czech Republic) and Observatorio del Montsec (Spain), respectively. We also wish to recognize the late Jaroslav Trnka, who observed a transit of TrES-1 b at Altan Observatory (Czech Republic). It is an honor to include his observation in this study.

Exoplanet Watch is a citizen science project managed by NASA’s Jet Propulsion Laboratory (JPL) on behalf of NASA’s Universe of Learning and is supported by NASA under award number NNX16AC65A to the Space Telescope Science Institute, in partnership with Caltech/IPAC, the Center for Astrophysics | Harvard \& Smithsonian, and the NASA Jet Propulsion Laboratory.

This research has made use of the NASA Exoplanet Archive, which is operated by the California Institute of Technology, under contract with the National Aeronautics and Space Administration under the Exoplanet Exploration Program. We also made use of the VizieR catalogue access tool, CDS, Strasbourg, France \citep{VizieR,Vizier2000}.

This paper includes data collected by the Transiting Exoplanet Survey Satellite (TESS; \citealt{Ricker2015}), which can be accessed via \dataset[https://doi.org/10.17909/v72t-xk76]{https://doi.org/10.17909/v72t-xk76}. These data are publicly available via the Mikulski Archive for Space Telescopes (MAST), and were processed by the Science Processing Operations Center (SPOC; \citealt{Jenkins2016}) at NASA Ames Research Center. Funding for the TESS mission is provided by NASA’s Science Mission Directorate.

This work further incorporated observations obtained with the Spitzer Space Telescope, which was operated by JPL, California Institute of Technology, under a contract with NASA. Those observations were conducted using the IRAC \citep{Fazio2004} and IRS \citep{Houck2004} instruments.

Finally, this research was supported in part by an NSERC Discovery Grant (DG-2020-04635) and the University of British Columbia. SH was supported in part by an NSERC Postgraduate Scholarship (PGS-D) and a Li Tze Fong Fellowship.

\facilities{TESS, HST, MAST, AAVSO, Spitzer, Exoplanet Archive}

\software{\texttt{\href{https://github.com/simonehagey/orbdot}{OrbDot}} \citep{orbdot}, \texttt{pylightcurve} \citep{pylightcurve}, \texttt{ExoTETHyS} \citep{exotethys}, \texttt{nestle} \citep{nestle}, \texttt{Astropy} \citep{astropyI,astropyII,astropyIII}, \texttt{emcee} \citep{emcee}, \texttt{Matplotlib} \citep{matplotlib}, \texttt{corner} \citep{corner}, \texttt{Numpy} \citep{numpy}, \texttt{SciPy} \citep{scipy}.}

\bibliography{references}{}
\bibliographystyle{aasjournal}

\newpage
\appendix

\twocolumngrid

\section{Generalized Orbital Decay Equations} \label{appendix:tides_general_equations}
This appendix presents the full equations governing orbital decay due to tidal dissipation, without assumptions about the planetary system's architecture. These equations remain exact for any values of eccentricity, obliquity, and spin frequency.

Prior studies typically express the decay rate in terms of the semi-major axis ($da/dt$), but for observational applications, it is more useful to frame them in terms of the period derivative ($dP/dt$). Our goal is to present the period decay rate equations in a form that is both exact and accessible, as deriving them from the literature can be nontrivial.

We adopt the constant time lag (or ``viscous'') equilibrium tide model \citep[e.g.,][]{Hut1981,Levrard2007,Correia2010,Leconte2010}, treating the contributions from the star and planet separately. Since this model is based on the quadrupole approximation of the tidal response, the stellar and planetary contributions can be computed independently and summed afterward \citep{Leconte2010}.

To avoid confusion with commonly used symbols for orbital elements, we have modified several variable names from \citet{Leconte2010}:
\[[\omega, N_a(e), N(e), \Omega(e)] \rightarrow [\dot{\theta}, f(e), g(e), h(e)].\]

\subsection{Tidal Dissipation in the Host Star}\label{appendix:star_tides}
Using Equations (2), (5), and (19) from \citet{Leconte2010}, along with the relation
\[\frac{dP}{dt} = \frac{3P}{2a} \frac{da}{dt},\]
we obtain the orbital period decay rate due to stellar tides:
\begin{equation} \label{eq:pdot_tides_star}
    \begin{aligned}
    \frac{dP}{dt}_{\rm (star)} = -\frac{27\pi}{Q_\star^{'}} &\left(\frac{M_p}{M_\star}\right) \left(\frac{R_\star}{a}\right)^5 \\
    & \times \left[f(e) - g(e) \cos{\varepsilon_\star} \frac{\dot{\theta}_\star}{n}\right].
    \end{aligned}
\end{equation}
Here, $Q_\star^{'}$ is the modified annual tidal quality factor of the star, $\varepsilon_\star$ is the stellar obliquity, defined as the angle between the stellar spin and orbital angular momentum vectors, and $\dot{\theta}_\star$ is the stellar spin frequency. The orbital mean motion is denoted by $n$, while $a$ is the semi-major axis. The masses of the planet and star are given by $M_p$ and $M_\star$, respectively, and $R_\star$ is the stellar radius. The functions $f(e)$ and $g(e)$, which depend on the orbital eccentricity $e$, are defined as
\begin{equation} \label{eq:tides_f}
    f(e) = \frac{1+\frac{31}{2} e^2+\frac{255}{8} e^4+\frac{185}{16} e^6+\frac{25}{64} e^8}{(1-e^2)^{15/2}},
\end{equation}
\begin{equation} \label{eq:tides_g}
    g(e) = \frac{1+\frac{15}{2} e^2+\frac{45}{8} e^4+\frac{5}{16} e^6}{(1-e^2)^6}.
\end{equation}

\subsection{Tidal Dissipation in the Planet}\label{appendix:planet_tides}
Following the same approach as for the star, the orbital period decay rate due to planetary tides is
\begin{equation*} \label{eq:pdot_tides_planet_general}
    \frac{dP}{dt}_{\rm (planet)} = -\frac{27\pi}{Q_p^{'}} \left(\frac{M_\star}{M_p}\right) \left(\frac{R_p}{a}\right)^5 \left[f(e) - g(e) \cos{\varepsilon_p} \frac{\dot{\theta}_p}{n}\right].
\end{equation*}

In this case, $Q_p^{'}$ is the modified annual tidal quality factor of the planet, and $\varepsilon_p$ is the planetary obliquity, defined as the angle between the planetary spin and orbital angular momentum vectors. The planetary spin frequency is given by $\dot{\theta}_p$, and the planetary radius is $R_p$. The functions $f(e)$ and $g(e)$ are the same as defined in Equations \ref{eq:tides_f} and \ref{eq:tides_g}.

Since planetary rotation is typically unconstrained, a reasonable assumption is that the planet's spin has evolved toward a tidal equilibrium rate that is dependent on the eccentricity and obliquity \citep{Leconte2010, Correia2010}, given by
\begin{equation*}
    \dot{\theta}_{eq} \equiv n \, \frac{g(e)}{h(e)}\frac{2\cos{\varepsilon_p}}{(1 + \cos^2{\varepsilon_p})},
\end{equation*}
where,
\begin{equation} \label{eq:tides_h}
    h(e) = \frac{1+3 e^2+\frac{3}{8} e^4}{(1-e^2)^{9/2}}.
\end{equation}

Substituting this equilibrium spin rate into the equation above gives
\begin{equation} \label{eq:pdot_tides_planet}
    \begin{aligned}
    \frac{dP}{dt}_{\rm (planet)} = -\frac{27\pi}{Q_p^{'}} &\left(\frac{M_\star}{M_p}\right) \left(\frac{R_p}{a}\right)^5 \\
    & \times \left[f(e) - \frac{g^2(e)}{h(e)}\frac{2 \cos^2{\varepsilon_p}}{(1 + \cos^2{\varepsilon_p})}\right],
    \end{aligned}
\end{equation}

This expression accounts for both eccentricity-driven and obliquity-driven tidal dissipation, providing a unified treatment of planetary tides.  

Equation \ref{eq:pdot_tides_planet} shows that if the orbit is circular and aligned, meaning $e = 0$ and $\varepsilon_p = 0$, the planetary contribution to orbital decay vanishes. In this case, the functions $f(e)$, $g(e)$, and $h(e)$ approach unity, and the equilibrium spin rate synchronizes with the mean motion, $\dot{\theta}_p = n$, leading to no net tidal energy dissipation in the planet.

\clearpage
\onecolumngrid

\section{Full Results Table}\label{appendix:results_table}


\begin{longtable}{|lcc||c|c|c|}
    \caption{Summary of orbital model fits to TrES-1\,b data, including constant period (circular and eccentric), orbital decay (circular and eccentric), and apsidal precession. Each model was tested with different data subsets: all data (transit, eclipse, RV), data without eclipses (transit, RV), and timing data only (transit, eclipse). Uncertainties represent 68\% credible intervals on the posteriors. RV jitter values were fixed based on a prior circular RV model fit. Models are ranked within each column by Bayesian evidence, with apsidal precession providing the best fit in all cases. Bayesian evidence should be compared only within columns, not across data subsets. To compare evidence values quantitatively, the Bayes factor should be calculated using Equation \ref{eq:bayes_factor} and assessed according to the thresholds in Table \ref{tab:bayesian_evidence} (see Section \ref{sec:results})} \label{tab:all_results} \\ \hline

    \textbf{Sym.} & \textbf{Unit} & \textbf{Prior} & \textbf{68\% Credible Interval} & \textbf{68\% Credible Interval} & \textbf{68\% Credible Interval} \\
    \hline\hline
    \endfirsthead

    \hline
    \textbf{Sym.} & \textbf{Unit} & \textbf{Prior} & \textbf{68\% Credible Interval} & \textbf{68\% Credible Interval} & \textbf{68\% Credible Interval} \\
    \hline\hline 
    \endhead

    \hline
    \multicolumn{6}{r}{\textit{Continued on next page...}} \\
    \endfoot

    \hline
    \endlastfoot

    \hline \noalign{\vskip 3mm}
    \multicolumn{3}{l}{\textbf{\normalsize{Constant Period -- Circular}}} & \multicolumn{1}{c}{\textbf{All Data}} & \multicolumn{1}{c}{\textbf{Without Eclipses}} & \multicolumn{1}{c}{\textbf{Only Mid-Times}} \\ \hline
    \rule{0pt}{2.7ex}$\log{\mathrm{Z}}$ & -- & -- & $-345.39$ & $-339.47$ & $-210.21$ \\[2pt] 
    $t_0$ & BJD$_{\mathrm{TDB}}$ & $\mathcal{N}(2456368.3807,0.01)$ & $2456368.380747\,^{+\,0.000022}_{-0.000022}$ & $2456368.380746\,^{+\,0.000022}_{-0.000023}$ & $2456368.380747\,^{+\,0.000022}_{-0.000022}$ \\[6pt]
    $P$ & days & $\mathcal{N}(3.03007,0.001)$ & $3.030069557\,^{+\,0.000000024}_{-0.000000024}$ & $3.030069557\,^{+\,0.000000024}_{-0.000000023}$ & $3.030069557\,^{+\,0.000000023}_{-0.000000026}$ \\[5pt]
    $K$ & m s$^{-1}$ & $\mathcal{U}(100,120)$ & $109.4\,^{+\,1.9}_{-1.9}$ & $109.3\,^{+\,1.9}_{-1.9}$ & -- \\[5pt]
    $\gamma_A$ & m s$^{-1}$ & $\mathcal{U}(-10,10)$ & $-2.1\,^{+\,4.4}_{-4.1}$ & $-2.1\,^{+\,4.5}_{-4.1}$ & -- \\[5pt] 
    $\gamma_B$ & m s$^{-1}$ & $\mathcal{U}(-20470,-20440)$ & $-20458.6\,^{+\,1.9}_{-1.8}$ & $-20458.6\,^{+\,2.0}_{-1.9}$ & -- \\[5pt]
    $\gamma_L$ & m s$^{-1}$ & $\mathcal{U}(-10,10)$ & $-5.3\,^{+\,1.5}_{-1.4}$ & $-5.3\,^{+\,1.5}_{-1.5}$ & -- \\[5pt]
    $\gamma_N$ & m s$^{-1}$ & $\mathcal{U}(-10,10)$ & $-0.2\,^{+\,3.9}_{-3.9}$ & $-0.3\,^{+\,3.9}_{-3.8}$  & -- \\[5pt]
    $\sigma_A$ & m s$^{-1}$ & Fixed & $0.34$ & $0.29$ & -- \\[2.6pt]
    $\sigma_B$ & m s$^{-1}$ & Fixed & $6.3$ & $6.4$ & -- \\[2.6pt]
    $\sigma_L$ & m s$^{-1}$ & Fixed & $0.39$ & $0.40$ & -- \\[2.6pt]
    $\sigma_N$ & m s$^{-1}$ & Fixed & $0.24$ & $0.24$ & -- \\[2.6pt]
    
    \hline \noalign{\vskip 5mm} 
    \multicolumn{3}{l}{\textbf{\normalsize{Constant Period -- Eccentric}}} & \multicolumn{1}{c}{\textbf{All Data}} & \multicolumn{1}{c}{} & \multicolumn{1}{c}{\textbf{Only Mid-Times}} \\ \hline
    \rule{0pt}{2.7ex}$\log{\mathrm{Z}}$ & -- & -- & $-342.15$ & -- & $-206.55$ \\[1pt] 
    $t_0$ & BJD$_{\mathrm{TDB}}$ & $\mathcal{N}(2456368.3807,0.01)$ & $2456368.380746\,^{+\,0.000023}_{-0.000022}$ & -- & $2456368.380745\,^{+\,0.000023}_{-0.000022}$ \\[6pt]
    $P$ & days & $\mathcal{N}(3.03007,0.001)$ & $3.030069558\,^{+\,0.000000024}_{-0.000000023}$ & -- & $3.030069557\,^{+\,0.000000025}_{-0.000000024}$ \\[6pt]
    $e$ & & $\mathcal{U}(0.0,0.01)$ & $0.00128\,^{+\,0.00037}_{-0.00037}$ & -- & $0.00128\,^{+\,0.00037}_{-0.00037}$ \\[5pt]
    $\omega$ & rad & Fixed & $0.0$ & -- & $0.0$ \\[5pt]
    $K$ & m s$^{-1}$ & $\mathcal{U}(100,120)$ & $109.2\,^{+\,1.9}_{-1.9}$ & -- & -- \\[5pt]
    $\gamma_A$ & m s$^{-1}$ & $\mathcal{U}(-10,10)$ & $-2.1\,^{+\,4.4}_{-4.0}$ & -- & -- \\[5pt]
    $\gamma_B$ & m s$^{-1}$ & $\mathcal{U}(-20470,-20440)$ & $-20458.5\,^{+\,1.8}_{-2.0}$ & -- & -- \\[5pt]
    $\gamma_L$ & m s$^{-1}$ & $\mathcal{U}(-10,10)$ & $-5.3\,^{+\,1.5}_{-1.5}$ & -- & -- \\[5pt]
    $\gamma_N$ & m s$^{-1}$ & $\mathcal{U}(-10,10)$ & $-0.5\,^{+\,3.9}_{-3.8}$ & --  & -- \\[5pt]
    $\sigma_A$ & m s$^{-1}$ & Fixed & $0.34$ & -- & -- \\[2.6pt]
    $\sigma_B$ & m s$^{-1}$ & Fixed & $6.3$ & -- & -- \\[2.6pt]
    $\sigma_L$ & m s$^{-1}$ & Fixed & $0.39$ & -- & -- \\[2.6pt]
    $\sigma_N$ & m s$^{-1}$ & Fixed & $0.24$ & -- & -- \\[2.6pt]
    
    \hline \noalign{\vskip 7mm} 

    \multicolumn{3}{l}{\textbf{\normalsize{Orbital Decay -- Circular}}} & \multicolumn{1}{c}{\textbf{All Data}} & \multicolumn{1}{c}{\textbf{Without Eclipses}} & \multicolumn{1}{c}{\textbf{Only Mid-Times}} \\ \hline
    \rule{0pt}{2.7ex}$\log{\mathrm{Z}}$ & -- & -- & $-340.15$ & $-333.95$ & $-204.57$ \\[1pt]
    $t_0$ & BJD$_{\mathrm{TDB}}$ & $\mathcal{N}(2456368.3807,0.01)$ & $2456368.381034\,^{+\,0.000065}_{-0.000063}$ & $2456368.381032\,^{+\,0.000065}_{-0.000062}$ & $2456368.381036\,^{+\,0.000064}_{-0.000062}$ \\[6pt]
    $P$ & days & $\mathcal{N}(3.03007,0.001)$ & $3.030069542\,^{+\,0.000000024}_{-0.000000025}$ & $3.030069542\,^{+\,0.000000024}_{-0.000000024}$ & $3.030069542\,^{+\,0.000000026}_{-0.000000024}$ \\[6pt]
    d$P/dE$ & days ${\rm E}^{-1}$ & $\mathcal{U}(-1\times10^{-7},1\times10^{-7})$ & $-6.9\times10^{-10}\,^{+\,1.5\times10^{-10}}_{-1.4\times10^{-10}}$ & $-6.9\times10^{-10}\,^{+\,1.4\times10^{-10}}_{-1.4\times10^{-10}}$ & $-6.9\times10^{-10}\,^{+\,1.4\times10^{-10}}_{-1.4\times10^{-10}}$ \\[6pt]
    $dP/dt$ & ms yr$^{-1}$ & Derived & $-7.2\,^{+\,1.5}_{-1.5}$ & $-7.1\,^{+\,1.5}_{-1.5}$ & $-7.2\,^{+\,1.5}_{-1.5}$ \\[5pt]
    $K$ & m s$^{-1}$ & $\mathcal{U}(100,120)$ & $109.3\,^{+\,1.8}_{-1.9}$ & $109.3\,^{+\,1.9}_{-1.8}$ & -- \\[5pt]
    $\gamma_A$ & m s$^{-1}$ & $\mathcal{U}(-10,10)$ & $-2.1\,^{+\,4.3}_{-4.1}$ & $-2.2\,^{+\,4.4}_{-3.9}$ & -- \\[5pt]
    $\gamma_B$ & m s$^{-1}$ & $\mathcal{U}(-20470,-20440)$ & $-20458.6\,^{+\,1.8}_{-1.8}$ & $-20458.5\,^{+\,1.9}_{-1.9}$ & -- \\[5pt]
    $\gamma_L$ & m s$^{-1}$ & $\mathcal{U}(-10,10)$ & $-5.3\,^{+\,1.5}_{-1.5}$ & $-5.3\,^{+\,1.5}_{-1.5}$ & -- \\[5pt]
    $\gamma_N$ & m s$^{-1}$ & $\mathcal{U}(-10,10)$ & $-0.3\,^{+\,4.0}_{-3.8}$ & $-0.3\,^{+\,3.7}_{-3.8}$  & -- \\[5pt]
    $\sigma_A$ & m s$^{-1}$ & Fixed & $0.34$ & $0.29$ & -- \\[2.6pt]
    $\sigma_B$ & m s$^{-1}$ & Fixed & $6.3$ & $6.4$ & -- \\[2.6pt]
    $\sigma_L$ & m s$^{-1}$ & Fixed & $0.39$ & $0.40$ & -- \\[2.6pt]
    $\sigma_N$ & m s$^{-1}$ & Fixed & $0.24$ & $0.24$ & -- \\[2.6pt]
    
    \hline \noalign{\vskip 6mm} 

    \multicolumn{3}{l}{\textbf{\normalsize{Orbital Decay -- Eccentric}}} & \multicolumn{1}{c}{\textbf{All Data}} & \multicolumn{1}{c}{} & \multicolumn{1}{c}{\textbf{Only Mid-Times}} \\ \hline
    \rule{0pt}{2.7ex}$\log{\mathrm{Z}}$ & -- & -- & $-336.43$ & -- & $-200.87$ \\[1pt]
    $t_0$ & BJD$_{\mathrm{TDB}}$ & $\mathcal{N}(2456368.3807,0.01)$ & $2456368.381033\,^{+\,0.000065}_{-0.000065}$ & -- & $2456368.381032\,^{+\,0.000064}_{-0.000062}$ \\[6pt]
    $P$ & days & $\mathcal{N}(3.03007,0.001)$ & $3.030069542\,^{+\,0.000000024}_{-0.000000024}$ & -- & $3.030069542\,^{+\,0.000000025}_{-0.000000025}$ \\[6pt]
    $dP/dE$ & days ${\rm E}^{-1}$ & $\mathcal{U}(-1\times10^{-7},1\times10^{-7})$ & $-6.8\times10^{-10}\,^{+\,1.4\times10^{-10}}_{-1.5\times10^{-10}}$ & -- & $-6.8\times10^{-10}\,^{+\,1.4\times10^{-10}}_{-1.4\times10^{-10}}$ \\[6pt]
    $dP/dt$ & ms yr$^{-1}$ & Derived & $-7.1\,^{+\,1.5}_{-1.6}$ & -- & $-7.1\,^{+\,1.4}_{-1.5}$ \\[6pt]
    $e$ & & $\mathcal{U}(0.0,0.01)$ & $0.00126\,^{+\,0.00036}_{-0.00037}$ & -- & $0.00126\,^{+\,0.00037}_{-0.00037}$ \\[5pt]
    $\omega$ & rad & Fixed & $0.0$ & -- & $0.0$ \\[5pt]
    $K$ & m s$^{-1}$ & $\mathcal{U}(100,120)$ & $109.2\,^{+\,1.9}_{-1.9}$ & -- & -- \\[5pt]
    $\gamma_A$ & m s$^{-1}$ & $\mathcal{U}(-10,10)$ & $-2.1\,^{+\,4.3}_{-4.0}$ & -- & -- \\[5pt]
    $\gamma_B$ & m s$^{-1}$ & $\mathcal{U}(-20470,-20440)$ & $-20458.5\,^{+\,1.9}_{-1.9}$ & -- & -- \\[5pt]
    $\gamma_L$ & m s$^{-1}$ & $\mathcal{U}(-10,10)$ & $-5.3\,^{+\,1.4}_{-1.6}$ & -- & -- \\[5pt]
    $\gamma_N$ & m s$^{-1}$ & $\mathcal{U}(-10,10)$ & $-0.4\,^{+\,3.8}_{-3.9}$ & -- & -- \\[5pt]
    $\sigma_A$ & m s$^{-1}$ & Fixed & $0.34$ & -- & -- \\[2.6pt]
    $\sigma_B$ & m s$^{-1}$ & Fixed & $6.3$ & -- & -- \\[2.6pt]
    $\sigma_L$ & m s$^{-1}$ & Fixed & $0.39$ & -- & -- \\[2.6pt]
    $\sigma_N$ & m s$^{-1}$ & Fixed & $0.24$ & -- & -- \\[2.6pt]
    
    \hline \noalign{\vskip 12mm} 
     
    \multicolumn{3}{l}{\textbf{\normalsize{Apsidal Precession}}} & \multicolumn{1}{c}{\textbf{All Data}} & \multicolumn{1}{c}{\textbf{Without Eclipses}} & \multicolumn{1}{c}{\textbf{Only Mid-Times}} \\ \hline
    \rule{0pt}{2.7ex}$\log{\mathrm{Z}}$ & -- & -- & $-334.58$ & $-329.76$ & $-199.69$ \\[1pt]
    $t_0$ & BJD$_{\mathrm{TDB}}$ & $\mathcal{N}(2456368.3807,0.01)$ & $2456368.37915\,^{+\,0.00072}_{-0.00065}$ & $2456368.3785\,^{+\,0.0013}_{-0.0023}$ & $2456368.37928\,^{+\,0.00065}_{-0.00073}$ \\[6pt]
    $P_s$ & days & $\mathcal{N}(3.03007,0.001)$ & $3.03006664\,^{+\,0.00000099}_{-0.00000083}$ & $3.0300698\,^{+\,0.0000011}_{-0.0000013}$ & $3.03006683\,^{+\,0.00000099}_{-0.00000093}$ \\[6pt]
    $e$ & & $\mathcal{U}(0.0,0.01)$ & $0.0059\,^{+\,0.0026}_{-0.0028}$ & $0.0035\,^{+\,0.0032}_{-0.0019}$ & $0.0054\,^{+\,0.0029}_{-0.0026}$ \\[5pt]
    $\omega_0$ & rad & $\mathcal{U}(\frac{\pi}{2},\frac{3\pi}{2})$ & $1.909\,^{+\,0.081}_{-0.052}$ & $3.33\,^{+\,0.69}_{-0.82}$ & $1.916\,^{+\,0.092}_{-0.059}$ \\[6pt]
    $d\omega/dE$ & rad E$^{-1}$ & $\mathcal{U}(0.0,0.001)$ & $0.00058\,^{+\,0.00016}_{-0.00009}$ & $0.00052\,^{+\,0.00024}_{-0.00015}$ & $0.00059\,^{+\,0.00018}_{-0.00009}$ \\[6pt]
    $d\omega/dt$ & deg yr$^{-1}$ & Derived & $4.0\,^{+\,1.1}_{-0.6}$ & $3.6\,^{+\,1.7}_{-1.0}$ & $4.1\,^{+\,1.2}_{-0.7}$ \\[5pt]
    $K$ & m s$^{-1}$ & $\mathcal{U}(100,120)$ & $109.2\,^{+\,1.8}_{-1.8}$ & $109.4\,^{+\,1.8}_{-1.8}$ & -- \\[5pt]
    $\gamma_A$ & m s$^{-1}$ & $\mathcal{U}(-10,10)$ & $-2.1\,^{+\,3.9}_{-3.8}$ & $-2.1\,^{+\,4.2}_{-3.7}$ & -- \\[5pt]
    $\gamma_B$ & m s$^{-1}$ & $\mathcal{U}(-20470,-20440)$ & $-20458.6\,^{+\,1.8}_{-1.8}$ & $-20458.5\,^{+\,1.8}_{-1.8}$ & -- \\[5pt]
    $\gamma_L$ & m s$^{-1}$ & $\mathcal{U}(-10,10)$ & $-5.4\,^{+\,1.5}_{-1.5}$ & $-5.3\,^{+\,1.4}_{-1.4}$ & -- \\[5pt]
    $\gamma_N$ & m s$^{-1}$ & $\mathcal{U}(-10,10)$ & $-0.4\,^{+\,3.6}_{-3.5}$ & $0.5\,^{+\,3.7}_{-3.7}$  & -- \\[5pt]
    $\sigma_A$ & m s$^{-1}$ & Fixed & $0.34$ & $0.29$ & -- \\[2.6pt]
    $\sigma_B$ & m s$^{-1}$ & Fixed & $6.3$ & $6.4$ & -- \\[2.6pt]
    $\sigma_L$ & m s$^{-1}$ & Fixed & $0.39$ & $0.40$ & -- \\[2.6pt]
    $\sigma_N$ & m s$^{-1}$ & Fixed & $0.24$ & $0.24$ & -- \\[2.6pt]\hline
\end{longtable}

\clearpage

\section{Eccentric O-C Plot}\label{appendix:oc_eccentric}


\begin{figure}[ht]
\centering
    \includegraphics[width=\textwidth]{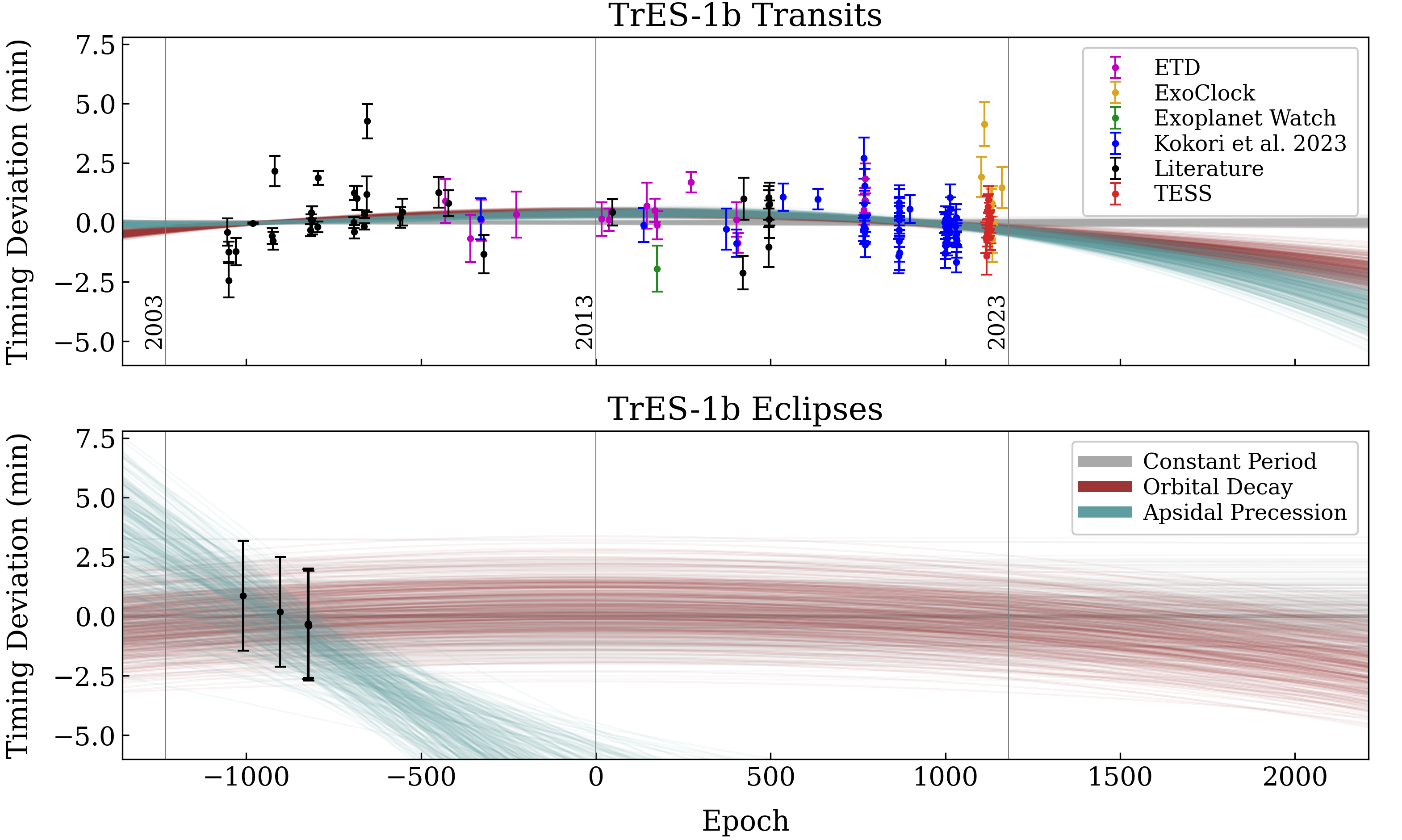}
\caption{Transit (top) and eclipse (bottom) timing variations for TrES-1 b, allowing for nonzero eccentricity in the constant-period and orbital decay models. The plot displays 300 random samples from the weighted posterior distributions of the joint model fits. Each data point represents the difference between an observed timing and the midpoint expected from the best-fitting constant-period, eccentric orbit model.}\label{fig:oc_eccentric}
\end{figure}

\clearpage

\section{Transit Timing Data}\label{appendix:transit_data}


\begin{longtable}{|c|c|c|c|c|c|}
    \caption{$^{1}$Kitt Peak National Observatory, USA. 
    $^{2}$Altan Observatory, Czech Republic. 
    $^{3}$Stonegate Observatory, USA. 
    $^{4}$Masaryk University Observatory, Czech Republic. 
    $^{5}$Private Observatory, Germany.  
    $^{6}$AO Prežganje, Slovenia. 
    $^{7}$Gothers Observatory, UK. 
    $^{8}$Quarryview Observatory/HQR. 
    $^{9}$POST, UK. 
    $^{10}$University of Siena Observatory, Italy. 
    $^{11}$Acton Sky Portal, USA. 
    $^{12}$UnderOak Observatory, USA. 
    $^{13}$Anunaki Observatory, Spain. 
    $^{14}$Observatoire des Baronnies Provençales, France. 
    $^{15}$Observatoire de Saint Véran, France. 
    $^{16}$Observatorio de Sencelles, Spain. 
    $^{17}$Observatoire de Vaison-La-Romaine, France. 
    $^{18}$Observatorio del Montsec, Spain. 
    $^{19}$Observatori Astronòmic Albanya, Spain. 
    $^{20}$Tartu Observatory, Estonia. 
    $^{21}$Private Observatory, France. 
    $^{22}$Private Observatory, UK. 
    $^{23}$Observatoire des Loges, France. 
    $^{24}$Burnham Observatory, UK. 
    $^{25}$Cavallino Observatory, Italy. 
    $^{26}$Observatoire du Guernet, France. 
    $^{27}$HRT Observatory, Spain. 
    $^{28}$Les Barres Observatory, France. 
    $^{29}$MarSEC, Italy. 
    $^{30}$TURMX Observatory, Spain. 
    $^{31}$Observatoire Privé du Mont, France. 
    $^{32}$Vierzon Observatory, France. 
    $^{33}$Private Observatory, Spain. 
    $^{34}$Thai Robotic Telescopes Sierra Remote Observatories, USA. 
    $^{35}$Observatoire des Valentines, Switzerland. 
    $^{36}$Blois Sologne Astronomie Observatory, France. 
    $^{37}$Private Observatory, Germany.} \label{tab:ttv_data} \\
    
    \hline
    \textbf{Epoch} & \textbf{Transit Midpoint (BJD$_\text{TDB}$)} & 
    \textbf{Error (days)} & \textbf{Source} & \textbf{Observer} & 
    \textbf{Re-fit?} \\
    \hline
    \endfirsthead
    
    \hline
    \textbf{Epoch} & \textbf{Transit Midpoint (BJD$_\text{TDB}$)} & 
    \textbf{Error (days)} & \textbf{Source} & \textbf{Observer} & 
    \textbf{Re-fit?} \\
    \hline
    \endhead
    
    \hline
    \multicolumn{6}{r}{\textit{Continued on next page...}} \\
    \endfoot
    
    \hline
    \endlastfoot
    
    $-1054$ & 2453174.68715 & 0.00040 & \citet{Charbonneau2005} & \citet{Alonso2004} & N \\
    $-1051$ & 2453183.77595 & 0.00050 & \citet{Charbonneau2005} & \citet{Alonso2004} & N \\
    $-1050$ & 2453186.80685 & 0.00030 & \citet{Charbonneau2005} & \citet{Alonso2004} & N \\
    $-1030$ & 2453247.40825 & 0.00040 & \citet{Charbonneau2005} & \citet{Alonso2004} & N \\
    $-981$ & 2453395.88249 & 0.00002 & \citet{rabus_t1_2009_hst} & HST & Y \\
    $-926$ & 2453562.53594 & 0.00023 & \citet{rabus_t1_2009} & --- & Y \\
    $-924$ & 2453568.59594 & 0.00026 & \citet{rabus_t1_2009} & --- & Y \\
    $-919$ & 2453583.74832 & 0.00044 & Private Communication & Pedro V. Sada$^{ 1}$ & Y \\
    $-816$ & 2453895.84376 & 0.00017 & \citet{Winn2007} & --- & Y \\
    $-815$ & 2453898.87415 & 0.00014 & \citet{Winn2007} & --- & Y \\
    $-814$ & 2453901.90442 & 0.00018 & \citet{Winn2007} & --- & Y \\
    $-812$ & 2453907.96432 & 0.00041 & \citet{Narita2007} & --- & Y \\
    $-796$ & 2453956.44525 & 0.00016 & \citet{rabus_t1_2009} & --- & Y \\
    $-795$ & 2453959.47675 & 0.00020 & \citet{rabus_t1_2009} & --- & Y \\
    $-693$ & 2454268.54254 & 0.00016 & \citet{rabus_t1_2009} & --- & Y \\
    $-692$ & 2454271.57234 & 0.00020 & \citet{rabus_t1_2009} & --- & Y \\
    $-691$ & 2454274.60354 & 0.00021 & \citet{rabus_t1_2009} & --- & Y \\
    $-684$ & 2454295.81388 & 0.00034 & Private Communication & Pedro V. Sada & Y \\
    $-663$ & 2454359.44451 & 0.00009 & Private Communication & \citet{Hrudkova2009} & Y \\
    $-662$ & 2454362.47494 & 0.00011 & Private Communication & \citet{Hrudkova2009} & Y \\
    $-655$ & 2454383.68601 & 0.00052 & \citet{Sada2012} & --- & Y \\
    $-654$ & 2454386.71821 & 0.00050 & \citet{Sada2012} & --- & Y \\
    $-560$ & 2454671.54194 & 0.00030 & \citet{rabus_t1_2009} & --- & Y \\
    $-554$ & 2454689.72251 & 0.00039 & Private Communication & Pedro V. Sada & Y \\
    $-450$ & 2455004.85032 & 0.00045 & Private Communication & Pedro V. Sada & Y \\
    $-431$ & 2455062.42139 & 0.00064 & ETD & Jaroslav Trnka$^{ 2}$ & Y \\
    $-422$ & 2455089.69195 & 0.00038 & ETD & Gary Vander Haagen$^{ 3}$ & Y \\ %
    $-359$ & 2455280.58531 & 0.00069 & ETD & Šárka Dyčková$^{ 4}$ & Y \\ %
    $-330$ & 2455368.45790 & 0.00060 & \cite{kokori_exoclockIII} & Manfred Raetz$^{ 5}$ & N \\
    $-321$ & 2455395.72749 & 0.00056 & Private Communication & Pedro V. Sada & Y \\
    $-329$ & 2455371.48793 & 0.00060 & ETD & Matej Mihelčič$^{ 6}$ & Y \\
    $-227$ & 2455680.55519 & 0.00067 & ETD & Darryl Sergison$^{ 7}$ & Y \\
    $17$ & 2456419.89203 & 0.00049 & ETD & Ken Hose$^{ 8}$ & Y \\
    $37$ & 2456480.49340 & 0.00033 & ETD & Mark Salisbury$^{ 9}$ & Y \\
    $47$ & 2456510.79431 & 0.00038 & Private Communication & Pedro V. Sada & Y\\
    $137$ & 2456783.50020 & 0.00050 & \cite{kokori_exoclockIII} & Alessandro Marchini$^{ 10}$ & N \\
    $146$ & 2456810.77139 & 0.00067 & ETD & Paul Benni$^{ 11}$ & Y \\
    $169$ & 2456880.46285 & 0.00034 & ETD & Mark Salisbury & Y \\
    $175$ & 2456898.64284 & 0.00041 & Exoplanet Watch & Kevin B. Alton$^{ 12}$ & Y \\
    $272$ & 2457192.56084 & 0.00030 & ETD & David Molina$^{ 13}$ & Y \\
    $373$ & 2457498.59650 & 0.00060 & \cite{kokori_exoclockIII} & Anaël Wünsche$^{ 14}$ & N \\
    $402$ & 2457586.46879 & 0.00050 & ETD & Marc Bretton$^{ 14}$ & Y \\
    $402$ & 2457586.46810 & 0.00040 & \cite{kokori_exoclockIII} & Anaël Wünsche & N \\
    $406$ & 2457598.58839 & 0.00029 & ETD & Jean-Christophe Dalouzy$^{ 15}$ & Y \\
    $420$ & 2457641.00849 & 0.00049 & \citet{Wang2021} & --- & Y \\
    $423$ & 2457650.10086 & 0.00061 & \citet{Wang2021} & --- & Y \\
    $494$ & 2457865.23439 & 0.00058 & \citet{Wang2021} & --- & Y \\
    $494$ & 2457865.23562 & 0.00043 & \citet{Wang2021} & --- & Y \\
    $495$ & 2457868.26591 & 0.00033 & \citet{Wang2021} & --- & Y \\
    $496$ & 2457871.29534 & 0.00055 & \citet{Wang2021} & --- & Y \\
    $497$ & 2457874.32585 & 0.00063 & \citet{Wang2021} & --- & Y \\
    $535$ & 2457989.46870 & 0.00040 & \cite{kokori_exoclockIII} & Mario Morales-Aimar$^{16}$ & N \\ 
    $635$ & 2458292.47560 & 0.00030 & \cite{kokori_exoclockIII} & Anaël Wünsche & N \\
    $765$ & 2458686.38375 & 0.00034 & \cite{kokori_exoclockIII} & TESS & N \\
    $766$ & 2458689.41393 & 0.00031 & \cite{kokori_exoclockIII} & TESS & N \\
    $766$ & 2458689.41590 & 0.00060 & \cite{kokori_exoclockIII} & Anaël Wünsche & N \\
    $766$ & 2458689.41440 & 0.00044 & ETD & Yves Jongen$^{17}$ & Y \\
    $768$ & 2458695.47471 & 0.00038 & \cite{kokori_exoclockIII} & TESS & N \\
    $769$ & 2458698.50488 & 0.00036 & ETD & Yves Jongen & Y \\
    $770$ & 2458701.53559 & 0.00044 & ETD & Vicenç Ferrando$^{ 18}$ & Y \\
    $769$ & 2458698.50395 & 0.00034 & \cite{kokori_exoclockIII} & TESS & N \\
    $769$ & 2458698.50530 & 0.00050 & \cite{kokori_exoclockIII} & Mario Morales-Aimar & N \\
    $771$ & 2458704.56373 & 0.00037 & \cite{kokori_exoclockIII} & TESS & N \\
    $771$ & 2458704.56420 & 0.00040 & \cite{kokori_exoclockIII} & Pere Guerra$^{ 19}$ & N \\
    $771$ & 2458704.56501 & 0.00038 & ETD & Yves Jongen & Y \\
    $866$ & 2458992.42000 & 0.00050 & \cite{kokori_exoclockIII} & Tõnis Eenmäe$^{ 20}$ & N \\
    $867$ & 2458995.45050 & 0.00060 & \cite{kokori_exoclockIII} & Jacques Michelet$^{ 21}$ & N \\
    $867$ & 2458995.45150 & 0.00030 & \cite{kokori_exoclockIII} & Stephane Ferratfiat$^{ 14}$ & N \\
    $868$ & 2458998.48120 & 0.00050 & \cite{kokori_exoclockIII} & Adrian Jones$^{ 22}$ & N \\
    $868$ & 2458998.48170 & 0.00050 & \cite{kokori_exoclockIII} & Anaël Wünsche & N \\
    $868$ & 2458998.48120 & 0.00040 & \cite{kokori_exoclockIII} & Emmanuel Besson$^{ 23}$ & N \\
    $868$ & 2458998.48170 & 0.00040 & \cite{kokori_exoclockIII} & Martin Valentine Crow$^{ 24}$ & N \\
    $868$ & 2458998.48130 & 0.00040 & \cite{kokori_exoclockIII} & Mauro Caló$^{ 25}$ & N \\
    $868$ & 2458998.48090 & 0.00050 & \cite{kokori_exoclockIII} & Patrick Brandebourg$^{ 26}$ & N \\
    $868$ & 2458998.48130 & 0.00040 & \cite{kokori_exoclockIII} & Pere Guerra & N \\
    $869$ & 2459001.51030 & 0.00050 & \cite{kokori_exoclockIII} & François Regembal$^{ 27}$ & N \\
    $869$ & 2459001.51150 & 0.00040 & \cite{kokori_exoclockIII} & Manfred Raetz & N \\
    $898$ & 2459089.38360 & 0.00040 & \cite{kokori_exoclockIII} & Yves Jongen & N \\
    $998$ & 2459392.39016 & 0.00031 & \cite{kokori_exoclockIII} & TESS & N \\
    $999$ & 2459395.41933 & 0.00042 & \cite{kokori_exoclockIII} & TESS & N \\
    $1000$ & 2459398.44970 & 0.00030 & \cite{kokori_exoclockIII} & Anaël Wünsche & N \\
    $1000$ & 2459398.45039 & 0.00039 & \cite{kokori_exoclockIII} & TESS & N \\
    $1001$ & 2459401.47970 & 0.00040 & \cite{kokori_exoclockIII} & Marc Deldem$^{ 28}$ & N \\
    $1001$ & 2459401.48025 & 0.00036 & \cite{kokori_exoclockIII} & TESS & N \\
    $1003$ & 2459407.54045 & 0.00032 & \cite{kokori_exoclockIII} & TESS & N \\
    $1005$ & 2459413.60031 & 0.00035 & \cite{kokori_exoclockIII} & TESS & N \\
    $1006$ & 2459416.63008 & 0.00032 & \cite{kokori_exoclockIII} & TESS & N \\
    $1008$ & 2459422.69075 & 0.00035 & \cite{kokori_exoclockIII} & TESS & N \\
    $1009$ & 2459425.72102 & 0.00035 & \cite{kokori_exoclockIII} & TESS & N \\
    $1011$ & 2459431.78083 & 0.00034 & \cite{kokori_exoclockIII} & TESS & N \\
    $1012$ & 2459434.81085 & 0.00035 & \cite{kokori_exoclockIII} & TESS & N \\
    $1013$ & 2459437.84194 & 0.00038 & \cite{kokori_exoclockIII} & TESS & N \\
    $1014$ & 2459440.87128 & 0.00035 & \cite{kokori_exoclockIII} & TESS & N \\
    $1015$ & 2459443.90174 & 0.00034 & \cite{kokori_exoclockIII} & TESS & N \\
    $1029$ & 2459486.32220 & 0.00050 & \cite{kokori_exoclockIII} & Alessandro Marchini & N \\
    $1030$ & 2459489.35200 & 0.00030 & \cite{kokori_exoclockIII} & Ivo Peretto$^{ 29}$ & N \\
    $1030$ & 2459489.35200 & 0.00030 & \cite{kokori_exoclockIII} & Mark Salisbury & N \\
    $1031$ & 2459492.38130 & 0.00030 & \cite{kokori_exoclockIII} & Robert Roth$^{ 30}$ & N \\
    $1031$ & 2459492.38199 & 0.00019 & \cite{kokori_exoclockIII} & Robert Roth & N \\
    $1031$ & 2459492.38260 & 0.00040 & \cite{kokori_exoclockIII} & Didier Laloum$^{ 31}$ & N \\
    $1032$ & 2459495.41200 & 0.00050 & \cite{kokori_exoclockIII} & Didier Laloum & N \\
    $1032$ & 2459495.41200 & 0.00050 & \cite{kokori_exoclockIII} & Lionel Rousselot$^{ 32}$ & N \\
    $1032$ & 2459495.41220 & 0.00040 & \cite{kokori_exoclockIII} & Robert Roth & N \\
    $1103$ & 2459710.54880 & 0.00059 & ExoClock & Miguel Ángel Álava Amat$^{ 33}$ & Y \\
    $1112$ & 2459737.82097 & 0.00065 & ExoClock & Napaporn A-thano$^{ 34}$ & Y \\
    $1115$ & 2459746.90786 & 0.00045 & TESS & --- & Y \\
    $1117$ & 2459752.96796 & 0.00040 & TESS & --- & Y \\
    $1118$ & 2459755.99753 & 0.00054 & TESS & --- & Y \\
    $1119$ & 2459759.02827 & 0.00039 & TESS & --- & Y \\
    $1121$ & 2459765.08903 & 0.00046 & TESS & --- & Y \\
    $1122$ & 2459768.11925 & 0.00036 & TESS & --- & Y \\
    $1123$ & 2459771.14953 & 0.00039 & TESS & --- & Y \\
    $1124$ & 2459774.17880 & 0.00044 & TESS & --- & Y \\
    $1125$ & 2459777.20919 & 0.00039 & TESS & --- & Y \\
    $1126$ & 2459780.23906 & 0.00038 & TESS & --- & Y \\
    $1128$ & 2459786.29876 & 0.00036 & TESS & --- & Y \\
    $1129$ & 2459789.32937 & 0.00039 & TESS & --- & Y \\
    $1130$ & 2459792.35913 & 0.00039 & TESS & --- & Y \\
    $1131$ & 2459795.38990 & 0.00049 & ExoClock & Stefano Lora$^{ 29}$ & Y \\
    $1133$ & 2459801.44908 & 0.00041 & ExoClock & Raphael Nicollerat$^{ 35}$ & Y \\
    $1133$ & 2459801.45012 & 0.00049 & ExoClock & Eric Miny$^{ 36}$ & Y \\
    $1134$ & 2459804.47913 & 0.00067 & ExoClock & Johannes Mieglitz$^{ 37}$ & Y \\
    $1162$ & 2459889.32259 & 0.00060 & ExoClock & Anaël Wünsche & Y \\
\end{longtable}
    \addtocounter{table}{-1}

\clearpage


\addtocounter{table}{+1}
\begin{longtable}{|c|c|c|c|c|}
\caption{VarAstro \citep{VarAstro} light curve ID numbers and observer names for TrES-1\,b transits from the Exoplanet Transit Database (ETD; \citet{poddany_etd_2010}). Transit mid-times and uncertainties are repeated here from Table \ref{tab:ttv_data} (in this Appendix) for convenience.\label{tab:etd_data}}\\
\hline
\textbf{VarAstro ID} & \textbf{Observer Name} & \textbf{Transit Midpoint (BJD$_\text{TDB}$)} & \textbf{Error (days)} & \textbf{Epoch} \\
\hline
\endfirsthead

\hline
\textbf{VarAstro ID} & \textbf{Observer Name} & \textbf{Transit Midpoint (BJD$_\text{TDB}$)} & \textbf{Error (days)} & \textbf{Epoch} \\
\hline
\endhead

\hline
\multicolumn{5}{r}{\textit{Continued on next page}} \\
\hline
\endfoot

\hline
\endlastfoot

82061 & Jaroslav Trnka & 2455062.42139 & 0.00064 & $-431$ \\
82063 & Gary Vander Haagen & 2455089.69195 & 0.00038 & $-422$ \\
82195 & Šárka Dyčková & 2455280.58531 & 0.00069 & $-359$ \\
82229 & Matej Mihelčič & 2455371.48793 & 0.00060 & $-329$ \\
82688 & Darryl Sergison & 2455680.55519 & 0.00067 & $-227$ \\
84041 & Ken Hose & 2456419.89203 & 0.00049 & $17$ \\
84161 & Mark Salisbury & 2456480.49340 & 0.00033 & $37$ \\
84856 & Paul Benni & 2456810.77139 & 0.00067 & $146$ \\
84944 & Mark Salisbury & 2456880.46285 & 0.00034 & $169$ \\
85476 & David Molina & 2457192.56084 & 0.00030 & $272$ \\
86112 & Marc Bretton & 2457586.46879 & 0.00050 & $402$ \\
86249 & Jean-Christophe Dalouzy & 2457598.58839 & 0.00029 & $406$ \\
88259 & Yves Jongen & 2458689.41440 & 0.00044 & $766$ \\
88282 & Yves Jongen & 2458698.50488 & 0.00036 & $769$ \\
88303 & Vicenç Ferrando & 2458701.53559 & 0.00044 & $770$ \\
88307 & Yves Jongen & 2458704.56501 & 0.00038 & $771$ \\[1pt]
\end{longtable}


\section{Eclipse Timing Data}\label{appendix:eclipse_data}



\begin{longtable}{|c|c|c|c|c|c|}
\caption{Secondary eclipse timing data for TrES-1\,b used in this study, based on Spitzer observations originally taken by \citet{Charbonneau2005} and reanalyzed by \citet{Cubillos2014}. The eclipse times are sourced directly from \citet{Cubillos2014} with a light travel time correction of $39.2$ seconds subtracted to account for the time delay across the orbit. The listed epochs correspond to the transit immediately preceding each eclipse.\label{tab:ttv_data_ecl}}\\
\hline
\textbf{Epoch} & \textbf{Eclipse Midpoint (BJD$_\text{TDB}$)} & \textbf{Error (days)} & \textbf{Source} & \textbf{Observer} & \textbf{Re-fit?} \\[2pt]
\hline
\endfirsthead

\hline
\textbf{Epoch} & \textbf{Eclipse Midpoint (BJD$_\text{TDB}$)} & \textbf{Error (days)} & \textbf{Source} & \textbf{Observer} & \textbf{Re-fit?} \\
\hline
\endhead

\hline
\multicolumn{6}{r}{\textit{Continued on next page}} \\
\hline
\endfoot

\hline
\endlastfoot

$-1010$ & 2453309.5286 & 0.0016 & \citet{Cubillos2014} & \citet{Charbonneau2005} & N \\[2pt]
$-904$  & 2453630.7155 & 0.0016 & \citet{Cubillos2014} & \citet{Charbonneau2005} & N \\[2pt]
$-824$  & 2453873.1207 & 0.0016 & \citet{Cubillos2014} & \citet{Charbonneau2005} & N \\[2pt]
$-823$  & 2453876.1508 & 0.0016 & \citet{Cubillos2014} & \citet{Charbonneau2005} & N \\[2pt]
$-822$  & 2453879.1808 & 0.0016 & \citet{Cubillos2014} & \citet{Charbonneau2005} & N \\[2.3pt] \hline
\end{longtable}


\clearpage

\section{Radial Velocity Data} \label{appendix:rv_data}



\begin{longtable}{|c|c|c|c|}
\caption{Radial velocity measurements of TrES-1 analyzed in this study.\label{tab:rv_data}}\\
\hline
\textbf{Time (BJD$_\text{TDB}$)} & \textbf{Velocity (m s$^{-1}$)} & \textbf{Error (m s$^{-1}$)} & \textbf{Data Source} \\
\hline
\endfirsthead

\hline
\textbf{Time (BJD$_\text{TDB}$)} & \textbf{Velocity (m s$^{-1}$)} & \textbf{Error (m s$^{-1}$)} & \textbf{Data Source} \\
\hline
\endhead

\hline
\multicolumn{4}{r}{\textit{Continued on next page}} \\
\hline
\endfoot

\hline
\endlastfoot

2453191.77075 & 60.4 & 12.8 & \citet{Alonso2004} \\
2453192.01275 & 115.1 & 8.3 & \citet{Alonso2004} \\
2453206.89175 & 87.1 & 16.0 & \citet{Alonso2004} \\
2453207.92675 & 15.8 & 10.4 & \citet{Alonso2004} \\
2453208.73075 & -113.3 & 15.0 & \citet{Alonso2004} \\
2453208.91775 & -98.1 & 19.8 & \citet{Alonso2004} \\
2453209.01875 & -118.4 & 15.3 & \citet{Alonso2004} \\
2453209.73175 & 49.8 & 15.7 & \citet{Alonso2004} \\
2453237.98000 & 68.32 & 3.66 & \citet{Laughlin2005} \\
2453238.84008 & -102.23 & 3.27 & \citet{Laughlin2005} \\
2453239.77435 & -24.53 & 3.25 & \citet{Laughlin2005} \\
2453239.88573 & 10.0 & 3.11 & \citet{Laughlin2005} \\
2453240.97760 & 70.68 & 3.73 & \citet{Laughlin2005} \\
2453907.87093 & 18.7 & 14.0 & \citet{Narita2007} \\
2453907.88214 & 30.5 & 12.5 & \citet{Narita2007} \\
2453907.90459 & 24.3 & 10.4 & \citet{Narita2007} \\
2453908.02807 & -17.7 & 13.6 & \citet{Narita2007} \\
2453908.03929 & -24.7 & 12.2 & \citet{Narita2007} \\
2453908.05053 & -27.5 & 11.1 & \citet{Narita2007} \\
2453908.06175 & -38.2 & 13.3 & \citet{Narita2007} \\
2453908.07298 & -23.7 & 11.2 & \citet{Narita2007} \\
2453908.08420 & -23.0 & 9.6 & \citet{Narita2007} \\
2456484.68861 & -20519.87 & 5.33 & \citet{Bonomo2017} \\
2456487.71076 & -20511.62 & 4.39 & \citet{Bonomo2017} \\
2456761.73339 & -20363.77 & 6.09 & \citet{Bonomo2017} \\
2456818.56378 & -20409.03 & 3.04 & \citet{Bonomo2017} \\
2456819.59861 & -20405.69 & 3.41 & \citet{Bonomo2017} \\
2456860.56373 & -20503.13 & 2.08 & \citet{Bonomo2017} \\
2456861.54517 & -20351.58 & 2.01 & \citet{Bonomo2017} \\
2456878.43180 & -20553.96 & 2.28 & \citet{Bonomo2017} \\
2457148.59138 & -20477.84 & 3.57 & \citet{Bonomo2017} \\
2457150.63238 & -20555.07 & 3.80 & \citet{Bonomo2017} \\
2457169.56345 & -20519.69 & 2.80 & \citet{Bonomo2017} \\
2457178.63558 & -20526.35 & 4.58 & \citet{Bonomo2017} \\
2457204.55116 & -20432.30 & 2.73 & \citet{Bonomo2017} \\
2457501.74874 & -20489.25 & 3.72 & \citet{Bonomo2017} \\
2457502.72711 & -20532.42 & 2.92 & \citet{Bonomo2017} \\
\end{longtable}


\end{document}